\documentclass[lettersize,journal]{IEEEtran}
\usepackage{array}
\usepackage[caption=false,font=normalsize,labelfont=sf,textfont=sf]{subfig}
\usepackage{textcomp}
\usepackage{stfloats}
\usepackage{url}
\usepackage{verbatim}
\usepackage{graphicx}
\usepackage{cite}
\usepackage{cite}
\usepackage{amsmath,amssymb,amsfonts}
\usepackage{graphicx}
\usepackage{textcomp}
\usepackage{multirow}
\usepackage{xcolor}
\usepackage[utf8]{inputenc}
\usepackage[bottom]{footmisc}
\usepackage{amsmath}
\usepackage{caption}
\usepackage{subcaption}
\usepackage{mathtools} 
\usepackage{tablefootnote}
\usepackage[euler]{textgreek}
\usepackage{tikz}
\usepackage{booktabs} 
\usepackage[utf8]{inputenc}
\usepackage{tcolorbox}
\usepackage[table]{xcolor}
\usepackage{enumitem}
\tcbuselibrary{listingsutf8}
\usepackage{minted} 
\definecolor{systemframe}{HTML}{558B2F}  
\definecolor{systemback}{HTML}{C5E1A5}  
\definecolor{userframe}{HTML}{558B2F}  
\definecolor{userback}{HTML}{C5E1A5}  
\usepackage{adjustbox}
\usepackage{graphicx}
\PassOptionsToPackage{most}{tcolorbox}
\usepackage{tcolorbox}
\usepackage{xcolor}
\usepackage[bookmarks=false, pdfencoding=auto, pdfborder={0 0 0}, pdfusetitle, draft]{hyperref}
\usepackage[most]{tcolorbox}
\usepackage{pifont}

\usepackage{siunitx}
\usepackage{multirow}
\usepackage{array}
\usepackage{verbatim}
\usepackage{stfloats}
\usepackage{url}
\usepackage{balance}
\usepackage{caption}
\usepackage{graphicx}
\usepackage{bbding}
\usepackage{pifont}
\usepackage{wasysym}
\usepackage{threeparttable}
\usepackage{longtable}
\usepackage{makecell}
\usepackage{algpseudocode}  
\usepackage{amsmath} 
\usepackage{threeparttable}
\usepackage{longtable}
\usepackage{booktabs}
\usepackage{makecell}
\usepackage{booktabs}
\usepackage{gensymb}
\usepackage[ruled]{algorithm2e}
\usepackage{tipa}%
\usepackage[commandnameprefix=always]{changes}
\newcommand{\systemname}{\text{Lip-Siri}}

\newcommand{\Rmnum}[1]{\uppercase\expandafter{\romannumeral #1}}

\begin{document}

\title{Lip-Siri: Contactless Open-Sentence Silent Speech with \\  Wi-Fi Backscatter}

\author{Ye~Tian,
        Haohua~Du, 
        Chao~Gu,
        Junyang~Zhang,
        Shanyue~Wang,
        Hao~Zhou,
        Jiahui~Hou,
        and~Xiang-Yang~Li,~\IEEEmembership{Fellow,~IEEE}%
        \thanks{Ye Tian, Chao Gu, Junyang Zhang, Shanyue Wang, Hao Zhou, Jiahui Hou, and Xiang-Yang Li were with the University of Science and Technology of China, Hefei, China. (e-mail: yetiancs@gmail.com) Haohua Du was with Beihang University, Beijing, China. (e-mail: duhaohua@buaa.edu.cn)}
}



\maketitle

\begin{abstract}
Silent speech interfaces (SSIs) enable silent interaction in noise-sensitive or privacy-sensitive settings. However, existing SSIs face practical deployment trade-offs among privacy, user experience, and energy consumption, and most remain limited to closed-set recognition over small, pre-defined vocabularies of words or sentences, which restricts real-world expressiveness. In this paper, we present {\systemname}, to the best of our knowledge, the first Wi-Fi backscatter--based SSI that supports open-vocabulary sentence recognition via lexicon-guided subword decoding. {\systemname} designs a frequency-shifted backscatter tag to isolate tag-modulated reflections and suppress interference from non-target motions, enabling reliable extraction of lip-motion traces from ubiquitous Wi-Fi signals. We then segment continuous traces into lip-motion units, cluster them, learn robust unit representations via cluster-based self-supervision, and finally propose a lexicon-guided Transformer encoder--decoder with beam search to decode variable-length sentence sequences. We implement an end-to-end prototype and evaluate it with 15 participants on 340 sentences and 3,398 words across multiple scenarios. {\systemname} achieves 85.61\% accuracy on word prediction and a WER of 36.87\% on continuous sentence recognition, approaching the performance of representative vision-based lip-reading systems.
\end{abstract}


\begin{figure*}[h]
  \centering
  \includegraphics[width=5.8in]{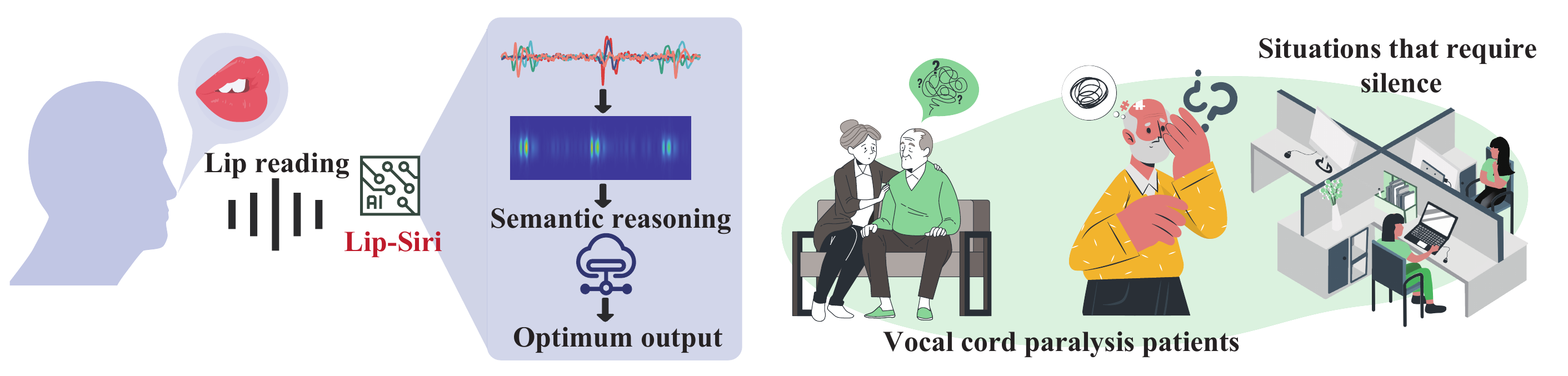} 
  \caption{The concept and use cases of {\systemname}: to provide non-contact and privacy protection silent lip-reading services for people, especially patients with speech disorders. It is also suitable for quiet situations that require silence or are inconvenient to speak, such as libraries, offices, dormitories, etc.}\label{first picture}
   \vspace{-0.2in}

\end{figure*}

\section{Introduction}

Speech communication is one of the most natural and efficient ways for people to interact with others. However, many individuals temporarily lose their voice due to vocal fold paralysis after surgery~\cite{aref2024vocal}, airway management~\cite{samuel2024stroboscopy}, or post-viral neuropathy~\cite{wang2025role}, and thus cannot communicate through normal spoken language. In clinical practice, vocal fold paralysis is a common complication. In the United States, approximately 20{,}000 new cases of unilateral vocal fold paralysis are diagnosed each year~\cite{lakpa2025job}. Most cases are reversible and involve only a transient loss of vocal function, but patients may suffer dysphonia or aphonia for weeks to months during recovery.
During this period, patients cannot produce clear speech, but they can make normal lip movements. For short-term patients, learning sign language is often impractical, motivating the development of a more efficient method that can interpret silent lip-reading to help them communicate.

In addition to assistive communication for patients, lip-reading recognition also enables silent speech interfaces (SSIs) for everyday interaction. It can serve as an enhanced interface in voice assistants for acoustically challenging environments (e.g., crowded streets or public transportation) where speech is easily corrupted by noise, as well as in quiet situations (e.g., libraries) where speaking aloud is inconvenient. Moreover, silently articulating commands can reduce unintended audio disclosure in public spaces, offering a more privacy-preserving interaction modality when users do not want nearby bystanders to overhear them. As a result, lip-reading recognition has attracted increasing attention in recent years~\cite{hameed2022pushing,ma2022visual,zhang2024lipwatch,akbar2025echolip}.

Early silent speech interfaces (SSIs) primarily relied on vision-based lip reading~\cite{assael2016lipnet,ma2022visual} or electromyographic (EMG) sensing~\cite{sugie1985speech,kapur2018alterego,kimura2022silentspeller}. Vision-based approaches can achieve strong accuracy under controlled conditions, but they are sensitive to lighting variations, occlusions (e.g., masks), and camera viewpoint, and video capture may be undesirable in privacy-sensitive scenarios. EMG-based approaches attach electrodes to the face or neck to measure muscle activity and infer articulation, but the required on-body instrumentation is intrusive and causes discomfort in long-term use.

Recently, researchers have explored sensing-based SSIs that infer lip motions from acoustic or inertial/strain signals~\cite{hameed2022pushing,wifihear}. Many of these systems are wearable-based: they leverage commodity devices such as smartphones~\cite{gao2020echowhisper,zhang2020endophasia,zhang2021soundlip,yin2023acoustic,akbar2025echolip}, earphones~\cite{jin2022earcommand}, smartwatches~\cite{zhang2024lipwatch}, glasses~\cite{zhang2023echospeech}, or VR headsets~\cite{zhang2021celip} to actively emit probing acoustic signals and analyze the echoes, or they use IMU/strain sensors to capture subtle facial vibrations~\cite{khanna2021jawsense,srivastava2022muteit,kunimi2022mask}. 
While these designs reduce the need for cameras, they assume that users are already wearing or holding specific devices. Moreover, they rely on continuous active probing (e.g., emitting ultrasound or audio chirps), which introduces practical constraints on battery-powered wearables. First, active probing competes with the device's primary I/O pipeline: for example, earphones may need to reserve the speaker and microphone chain for periodic probing and echo capture, which can conflict with normal audio playback and degrade user experience. 
Second, active probing incurs substantial energy overhead on power-constrained wearables, limiting long-term use.
Among RF-based SSIs, RFID Tattoo~\cite{wang2019rfid} and HearMe~\cite{zhang2022hearme} can sense lip motions from wireless reflections, but they require a tag to be attached or held near the user’s chin. In contrast, radar-based systems such as mSilent~\cite{zeng2023msilent} and TWLip-Seq~\cite{zhu2025twlip} achieve fully contactless lip reading by leveraging high-resolution radar sensing and deep models, yet expensive dedicated radar hardware limits ubiquitous deployment. Moreover, current sensing-based SSIs are still evaluated in a closed-set setting with a pre-defined vocabulary of words or sentences, restricting interaction flexibility and natural communication.




In this paper, we present {\systemname}, a contactless SSI that senses lip motions using Wi-Fi backscatter signals available in everyday indoor environments. {\systemname} employs low-power backscatter tags to modulate and reflect ambient Wi-Fi transmissions, while a receiver extracts lip-motion traces from the tag-shifted component for decoding. Because backscatter tags can operate at microwatt-level power~\cite{kellogg2016passive} and are inexpensive (on the order of a few dollars), the system offers a lightweight and cost-effective deployment option. 
More importantly, {\systemname} moves beyond closed-set command recognition by performing lexicon-guided subword sequence decoding. It constructs an extensible subword lexicon and decodes variable-length token sequences, enabling open-sentence recognition for any sentence that can be composed from the lexicon. This substantially expands the interaction space and better supports practical silent communication needs.

Designing and implementing {\systemname} poses several challenges. 
\textit{(a) Micro-motion sensing under interference.} Lip articulation induces subtle micro-movements around the lips and jaw, which are easily overwhelmed by non-target motions (e.g., head shifts, breathing, or nearby movers). Reliably isolating lip-motion signatures from Wi-Fi backscatter measurements is therefore non-trivial. 
\textit{(b) Robust feature learning under articulation variability.} 
Lip-motion patterns vary substantially across speakers; even for the same user, articulation speed and coarticulation differ across contexts. This variability makes it difficult to extract stable, generalizable representations for reliable decoding.
\textit{(c) Open-sentence decoding from continuous signals.} 
Unlike closed-set command classification, {\systemname} must map a continuous signal stream to a variable-length text sequence, which is very difficult. Because unit boundaries and frame-to-token alignments are unknown, and it requires context-aware decoding over a much larger open-sentence output space.

To address these challenges, we first conduct a preliminary study on 48 IPA phonemes and develop a motion-based signal model to understand how silent articulation modulates the Doppler characteristics of the backscattered signal, which guides robust feature design. Second, we design a frequency-shifted backscatter tag to isolate the tag-modulated component and suppress external interference and non-target body motions, yielding a clean lip-motion trace. Third, we approximately segment continuous traces into lip-motion units, cluster similar units, and perform cluster-based self-supervised pretraining to learn robust representations for lip-motion units under contextual variations. Finally, we enable open-sentence recognition via lexicon-guided subword decoding: we build an extensible subword lexicon and train a Transformer encoder--decoder to map lip-motion features to subword-token sequences, with beam search used to produce the most likely output under contextual constraints.

We invite 15 volunteers to evaluate {\systemname} on 340 sentences with 3,398 words across multiple scenarios.
Extensive experiments demonstrate the good performance of {\systemname}.
In summary, our main contributions are as follows:
\begin{itemize}
   \item We present {\systemname}, the first Wi-Fi backscatter--based SSI that enables open-sentence recognition via an extensible lexicon, rather than a closed set of pre-defined sentences.
    
    \item We design a frequency-shifted backscatter sensing and signal processing pipeline to isolate lip-motion signatures from Wi-Fi signals and suppress interference from external movers and non-target body motions.
    
    \item We propose a learning-and-decoding framework for continuous silent sentences. It captures lip-motion units via segmentation and clustering, pretrains robust representations with cluster-based self-supervision, and performs lexicon-guided sequence decoding with a Transformer encoder--decoder and beam search.
    
    \item We build the prototype and conduct extensive experiments. {\systemname} achieves 85.61\% accuracy on word prediction and 36.87\% WER on continuous sentence recognition. As the first sensing-based SSI for open-sentence decoding, it attains performance approaching that of a representative vision-based SSI while remaining effective in scenarios where cameras are unreliable.
\end{itemize}

\section{Related work}
\textbf{SSI based on Visual Signals.}
Visual-based SSI methods~\cite{potamianos1997speaker,chu2000bimodal,chung2017lip,son2017lip,shi2022learning} extract and interpret lip-reading features from images or videos.
Early studies typically relied on hand-crafted cues, such as color differences~\cite{wark1998approach,skodras2011unconstrained} and facial geometry/structures~\cite{fan2012system,puviarasan2011lip}, to localize the lip region and design the feature space.
With the rise of deep learning, researchers have developed end-to-end models that learn discriminative lip-motion representations directly from visual inputs.
For example, LipNet~\cite{assael2016lipnet} demonstrated sentence-level sequence prediction on constrained datasets.
Chung et al. further constructed large-scale datasets such as LRS2~\cite{afouras2018deep}, which contains over 1,000 speakers and a large vocabulary, enabling continuous lip-reading in the wild.
Building on these datasets, a series of studies~\cite{petridis2018end,ma2022visual,ma2023auto} have advanced continuous visual lip reading and reported WERs in the range of 33.6\% under standard settings.
Despite their strong performance, visual lip-reading approaches are sensitive to illumination and occlusion, and they become unreliable in low-light environments or when the speaker wears a mask.
Moreover, users may be concerned about video-privacy leakage, and recording videos can be inconvenient or not allowed in many daily scenarios (e.g., public spaces or privacy-sensitive environments).

\textbf{SSI based on EMG Signals.} EMG-based SSI \cite{janke2017emg,eskes2017predicting,dong2023electromyogram} infer lip-reading actions by measuring signal changes caused by muscle activity through electrodes placed on the user's face, neck, and oral cavity. Sugie and Tsunoda \cite{sugie1985speech} pioneered EMG-based SSI by recognizing five Japanese vowels using a three-channel electromyogram. Maier-Hein et al. \cite{maier2005session} and the AlterEgo project \cite{kapur2018alterego} achieved 97.3\% and 81\% accuracy, in recognizing digits and short phrases through EMG signals. Recently, SilentSpeller \cite{kimura2022silentspeller} designed a dental retainer to track tongue movements and generate EMG signals for recognizing spelled letters. These studies holds promise to help identify critical information expressed by voiceless patients.
However, EMG-based methods are highly sensitive to electrode placement, and positional deviations can significantly impact performance. Additionally, wearing electrode sensors can cause discomfort, particularly when inserted into the oral cavity or subcutaneously.

\textbf{SSI based on Sensing Signals.} 
Beyond vision and EMG, a growing body of work explores sensing-based SSIs that infer silent speech from non-acoustic signals, aiming to enable camera-free and potentially more privacy-friendly interaction. 
Wang et al.~\cite{wifihear} first demonstrated that Wi-Fi signals can capture users' mouth movements, while RFID-based designs such as RFID Tattoo~\cite{wang2019rfid} and HearMe~\cite{zhang2022hearme} exploit tag reflections for lip-motion recognition. More recently, radar-based systems (e.g., mSilent~\cite{zeng2023msilent} and TWLip-Seq~\cite{zhu2025twlip}) combine high-resolution sensing with deep models to improve detection performance. In parallel, some SSIs leverage sensors on devices that users wear or hold, such as IMU-based approaches (e.g., JawSense~\cite{khanna2021jawsense} and MuteIt~\cite{srivastava2022muteit}) and acoustic probing with commodity smartphones~\cite{tan2017silenttalk,zhang2020endophasia,zhang2021soundlip,yin2023acoustic,akbar2025echolip} or earphones~\cite{jin2022earcommand,sun2024earssr}. Despite these advances, current sensing-based SSIs still restrict their outputs to a fixed, pre-defined set of words or sentences, which limits expressiveness in real-world silent interaction. In contrast, {\systemname} adopts Wi-Fi backscatter sensing without requiring users to wear or hold any device, and enables open-sentence recognition via lexicon-guided decoding over an extensible lexicon, supporting previously undefined sentence compositions within the lexicon.

\section{Study the pattern of lip-reading movement}\label{sec:prestudy}

\subsection{Pre-experiment and Observation}


Speech is produced by air from the lungs vibrating the vocal cords, with sound quality and pitch shaped by resonating cavities like the mouth and nasal cavities. These modulations distinguish different phonemes. However, in silent lip-reading, vocal cord vibrations and cavity adjustments are absent. Wireless signals can detect lip-reading changes are mainly limited to in the facial area.
As shown in Fig.\ref{fig:Pre-study}(a) and Fig.\ref{fig:Pre-study}(b), during the process of lip reading, the mouth moves the surrounding muscles. Common mouth movements include closed lips, flat lips, biting the lower lip, pouting, naturally open, and widely open mouth. These different changes in motion have different effects on the signal.
To understand how silent articulation patterns manifest in our signals, we conduct a pilot study with two volunteers who silently articulated 48 English IPA phonemes. 
After preprocessing, we compare the signal patterns across phonemes and obtained key observations:


\textbf{(a) Vowels exhibit more pronounced signal variations than consonants.}
Vowels typically involve larger mouth opening and lip-shape changes; some diphthongs further introduce continuous transitions, resulting in stronger and more distinctive waveform variations (e.g., /\textipa{aI}/, /\textipa{OI}/, and /\textipa{aU}/).
\textbf{(b) Many consonants produce weak or indistinguishable signatures.}
For a subset of consonants, the mouth opening and visible lip motion are minimal during silent articulation (e.g., /m/, /\textipa{N}/, /ts/, and /dz/), leading to low-amplitude or unstable signal changes. 
This is a significant factor limiting the accuracy of lip-reading recognition, as much information is lost. 
It also suggests that inferring lip reading meanings by establishing phoneme combination rules might not be appropriate, because these missing consonant phonemes undermine the original grammatical rules.
\textbf{(c) Visually similar phonemes can be highly ambiguous in wireless signals.}
Some phoneme pairs share very similar mouth gestures (e.g., /s/ vs. /z/), whose difference in voiced speech mainly comes from vocal-fold vibration. 
Since wireless sensing primarily reflects mouth-shape dynamics but not voicing, their time-domain waveforms can overlap substantially, making them difficult to separate. Thus, how to distinguish these phonemes with similar mouth shapes is also the key to accurately identify the meaning of lip reading movements.

\begin{figure}[h]
 \centering
 \subfloat[\quad Some vowel.]{\includegraphics[height=1.4in]{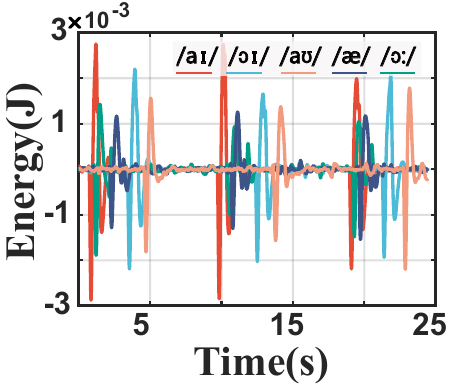}}
 \subfloat[\quad Some consonant.]{\includegraphics[height=1.4in]{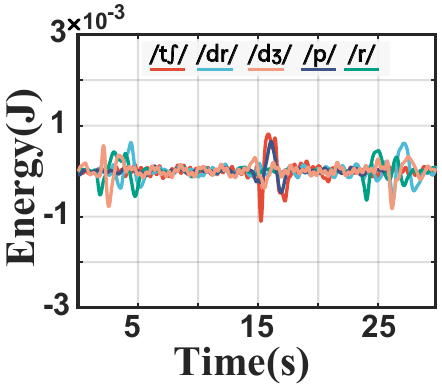}}
 \caption{ Comparison of normalized short-time energy curves for example vowel and consonant silent-articulation signals. Negative values indicate energy below the baseline due to normalization.}\label{y_and_f_compare}
\end{figure}

\begin{figure*}[t]
 \centering
 \subfloat[]{\includegraphics[height=2in]{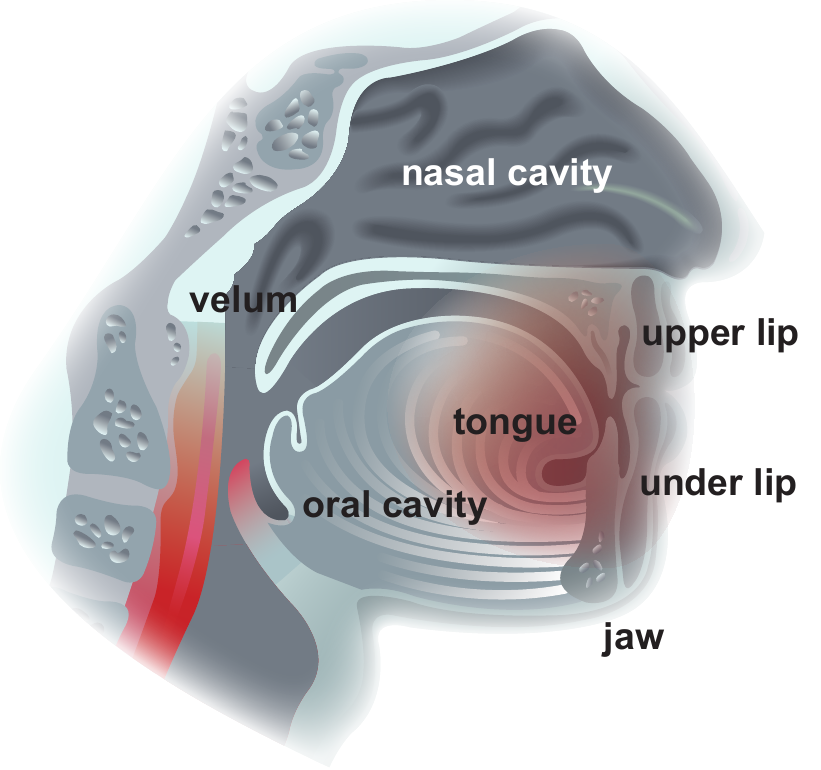}}
 \subfloat[]{\includegraphics[height=2in]{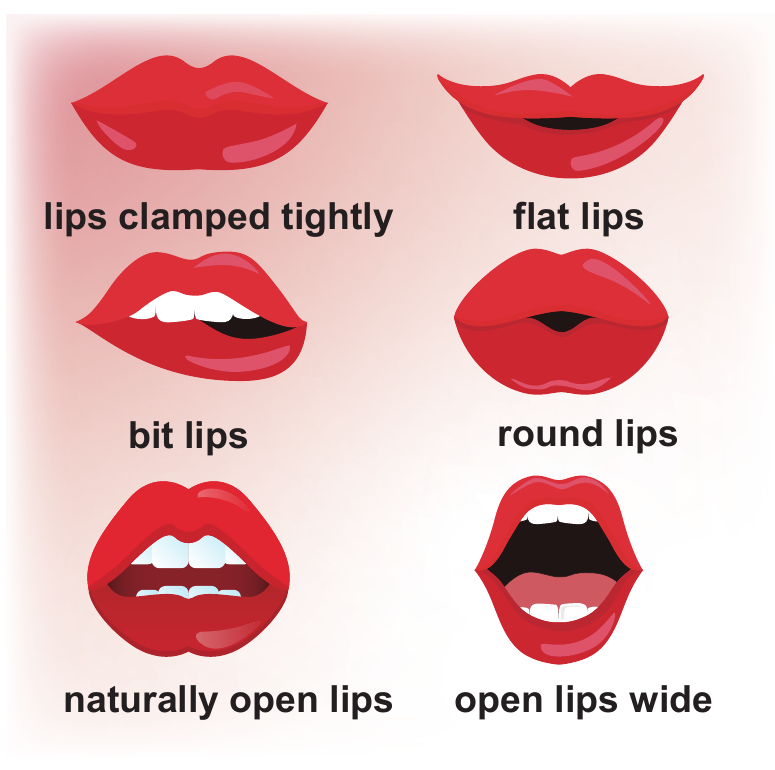}}
 \subfloat[]{\includegraphics[height=2in]{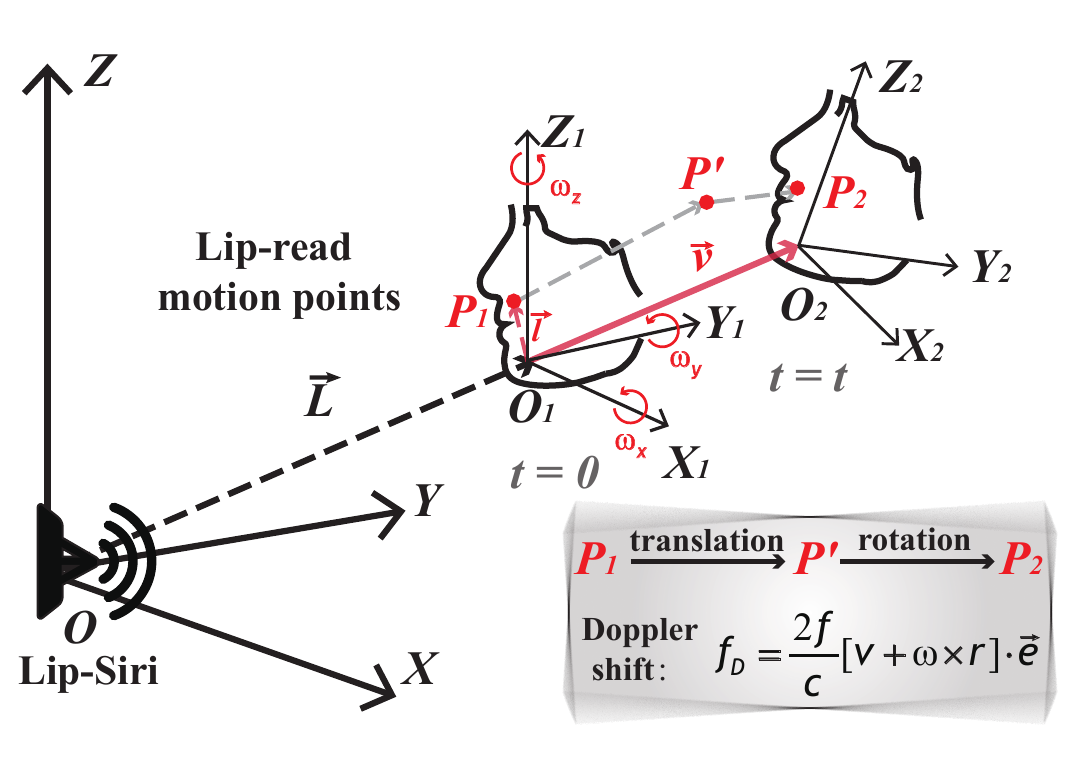}}
 \caption{Illustration and theoretical model of silent articulation for wireless sensing. 
(a) Key articulators and the main active region during silent articulation. 
(b) Representative mouth shapes that produce distinct lip-motion patterns. 
(c) Geometric motion model of lip movements and the resulting Doppler variation in the reflected signal.
}\label{fig:Pre-study}
\end{figure*}

\subsection{Theoretical Modeling and Analysis}
To better understand how silent articulation modulates wireless signals and to guide feature extraction, we build a motion-based model for lip-reading movements. The received signal can be viewed as the superposition of contributions from many moving scattering points on the face (dominated by the mouth and jaw region). Let $r(t)$ denote the received complex baseband signal:
\begin{equation}\label{eq:baseband}
r(t)=\sum_{i=1}^{N} a_i(t)\,e^{j\phi_i(t)} + n(t),
\end{equation}
where $a_i(t)$ and $\phi_i(t)$ represent the time-varying amplitude and phase contributed by the $i$-th scattering point, and $n(t)$ is additive noise. Intuitively, different mouth-shape transitions change both the geometry (relative positions) and reflectivity of the mouth region, leading to variations in signal intensity as well as the phase evolution of $r(t)$.

In practice, we characterize intensity variation using a short-time energy curve computed from the baseband magnitude:
\begin{equation}\label{eq:ste}
E(t)=\sum_{k=t}^{t+W-1} |r[k]|^2,
\end{equation}
with a sliding window of length $W$. We then extract statistics such as maximum/minimum/mean, standard deviation, and quantiles from $E(t)$ within each segment. Beyond average intensity, distributional statistics are also informative because the energy envelope is often asymmetric during continuous mouth-shape transitions. For example, some diphthongs exhibit different rising/falling dynamics when the mouth moves from a more open state to a narrower state (or vice versa), which can be reflected in the skewness and steepness of the energy distribution. (Note that when $E(t)$ is mean-centered or normalized for visualization, negative values indicate energy below a baseline rather than negative physical energy.)

\begin{figure}[t]
  \centering
  \includegraphics[width=2.8in]{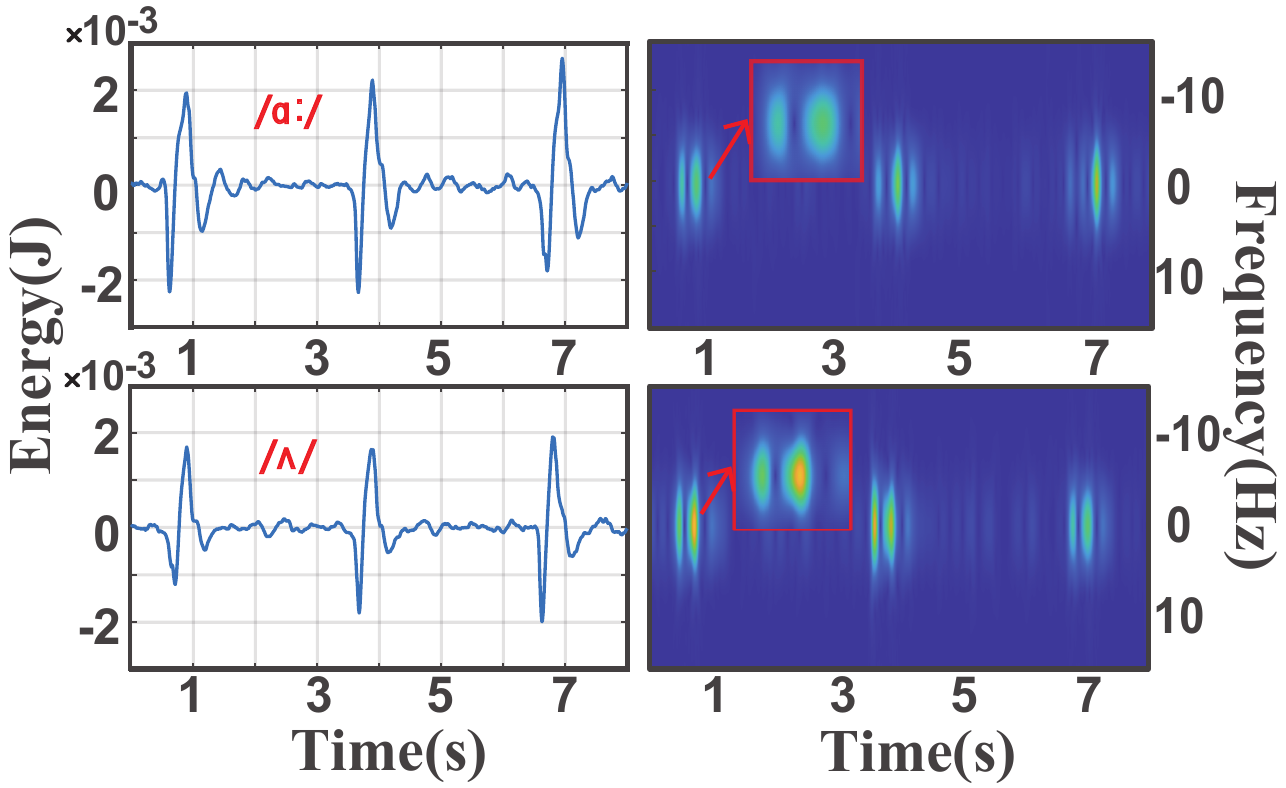} 
  \caption{Phonemes /a:/ and /\textturnv/. }\label{fig:aanda_L_compare}
\end{figure}

Frequency variation is mainly caused by Doppler shifts induced by mouth motions. As illustrated in Fig.~3(c), the motion of a scattering point on the mouth surface can be decomposed into translation and rotation. Consider a point $P$ whose initial position at $t=0$ is $P_1$. Let $O$ denote the transmit origin and let $O_1$ be the rotation center of a local mouth coordinate system. Define $\mathbf{L}=\overrightarrow{OO_1}$ as the vector from $O$ to $O_1$, and $\mathbf{l}=\overrightarrow{O_1P_1}$ as the position of $P_1$ in the local coordinate system. If the local coordinate system translates with velocity $\mathbf{v}$ and rotates with angular velocity $\boldsymbol{\omega}=(\omega_x,\omega_y,\omega_z)^{T}$, then the position of the point at time $t$ can be expressed as:
\begin{equation}\label{eq:pos}
\overrightarrow{OP(t)} = \mathbf{L} + \mathbf{v}t + \mathbf{R}(t)\mathbf{l},
\end{equation}
where $\mathbf{R}(t)$ is the time-varying rotation matrix. Differentiating Eq.~(\ref{eq:pos}) yields the instantaneous velocity:
\begin{equation}\label{eq:vel}
\dot{\overrightarrow{OP(t)}} = \mathbf{v} + \boldsymbol{\omega}\times(\mathbf{R}(t)\mathbf{l}),
\end{equation}
which contains both translational and rotational components. Since the instantaneous frequency equals the phase derivative, $f_D(t)=\frac{1}{2\pi}\frac{d\phi(t)}{dt}$, the Doppler shift is determined by the radial component of this velocity along the propagation direction. Under a monostatic approximation (equivalently, assuming the incident and reflected directions are similar over the small mouth displacement), the Doppler shift can be written as:
\begin{equation}\label{eq:doppler_simple}
f_D(t)=\frac{2f}{c}\Big(\mathbf{v}+\boldsymbol{\omega}\times\big(\mathbf{R}(t)\mathbf{l}\big)\Big)\cdot \hat{\mathbf{e}},
\end{equation}
where $f$ is the carrier frequency, $c$ is the wave speed, and $\hat{\mathbf{e}}$ is the unit vector along the propagation direction from the transmit and receive to the mouth region. Therefore, faster articulatory motions (larger $|\mathbf{v}|$ and/or $|\boldsymbol{\omega}|$) lead to more pronounced Doppler (frequency) variations. This also implies that some visemes with similar mouth shapes may still exhibit distinguishable frequency-domain signatures due to differences in articulation speed. To validate this, we compared several phonemes with highly similar mouth shapes and computed their spectrograms using STFT. For example, as shown in Fig.~\ref{fig:aanda_L_compare}, the time-domain waveforms of /a:/ and /\textturnv/ are highly similar, while /\textturnv/ exhibits faster mouth motions and thus more noticeable frequency-domain changes.

These analyses provide guidance for designing the core modules of {\systemname}. Specifically, since some phoneme-level cues may be weak or ambiguous in wireless observations, we avoid relying on hand-crafted phoneme-combination rules and instead adopt a context-dependent sequence prediction model. Guided by the above observations, we extract 67-dimensional features from both time- and frequency-domain cues, cluster signal-similar lip-reading syllable units, and employ self-supervised learning to capture the characteristics and variations of these similar units, which improves robustness under missing or overlapping low-level sensing features.

\begin{figure*}[t]
  \centering
  \includegraphics[width=6.6in]{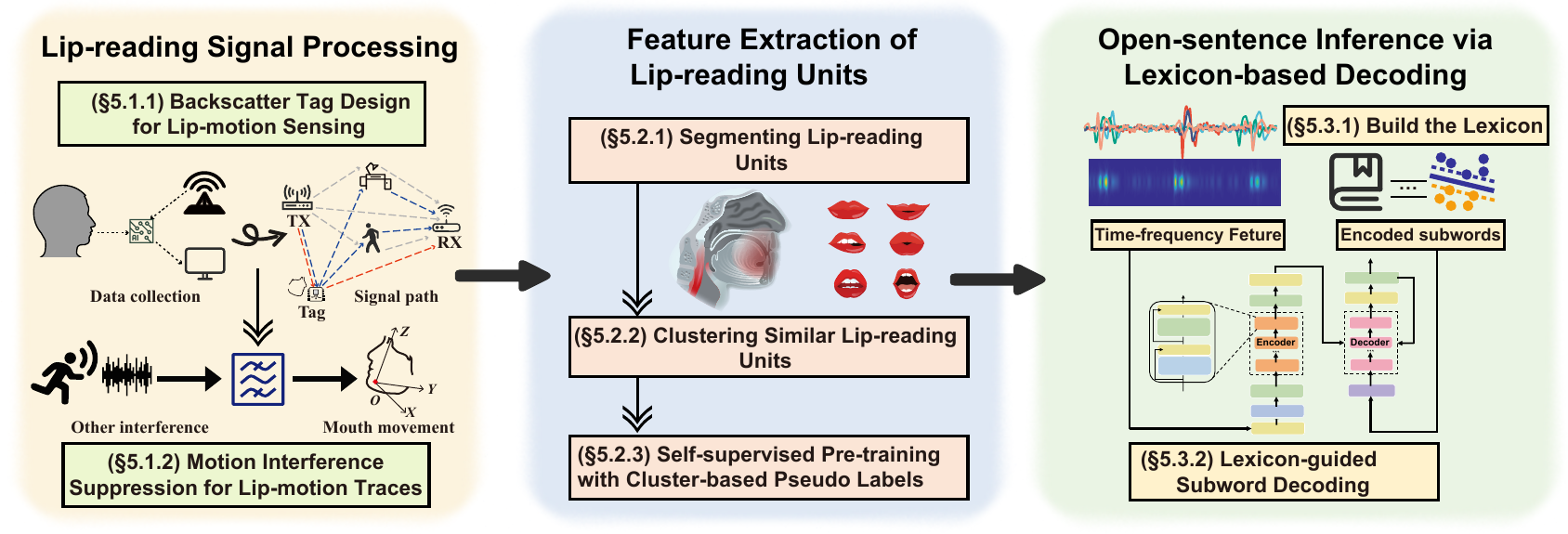} 
  \caption{
  System overview of {\systemname}: {\systemname} first extracts lip-motion traces from Wi-Fi backscatter signals and suppresses motion interference, then learns features of lip-motion units via segmentation, clustering, and self-supervised pretraining, and finally performs lexicon-based subword sequence decoding to enable open-sentence silent speech inference.}\label{fig:system overview}
\end{figure*}

\section{System Overview}

{\systemname} infers a user's silent utterance by sensing subtle lip motions with Wi-Fi backscatter signals. As illustrated in Fig.~\ref{fig:system overview}, the system consists of three core modules:
\textit{\textbf{(a) Lip-motion signal processing:} }
{\systemname} employs frequency-shifted backscatter sensing to isolate tag-modulated reflections that carry lip-motion signatures, and then suppresses motion interference from the environment and non-target body movements to obtain a clean lip-motion trace (Sec.~\ref{sec:signal processing}).
\textit{\textbf{(b) Lip-motion unit feature extraction:} }
From continuous lip-motion traces, {\systemname} approximately segments candidate lip-motion units, clusters similar units to form unit categories, and performs cluster-based self-supervised pretraining to learn robust unit representations that generalize across speakers and contexts (Sec.~\ref{sec:Feature Extraction}).
\textit{\textbf{(c) Open-sentence inference via lexicon-based decoding:}}
Building on the learned unit-level representations, {\systemname} constructs a subword lexicon and performs lexicon-guided sequence decoding with a Transformer encoder--decoder to generate subword-token sequences, enabling open-sentence recognition within the lexicon (Sec.~\ref{sec:Semantic Reasoning}).

\section{System Design}\label{sec:Design details}
\subsection{Lip-reading Signal Processing}\label{sec:signal processing}

\subsubsection{Backscatter Tag Design for Lip-motion Sensing}
\begin{figure}[h]
  \centering
  \includegraphics[width=2.3in]{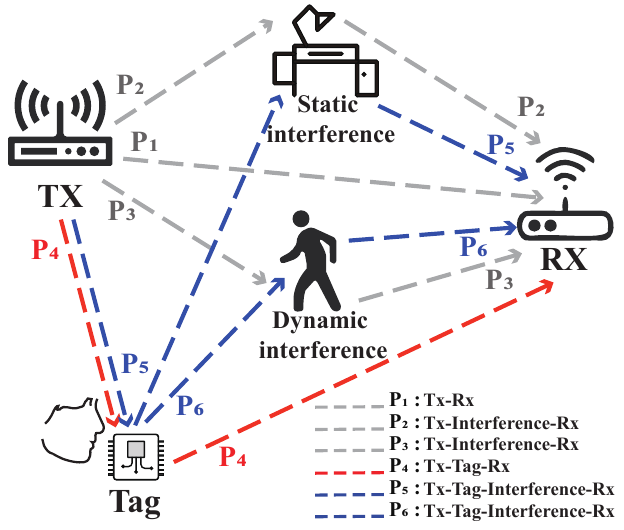} 
  \caption{Signal propagation paths in {\systemname}, including the direct path, tag-reflected path, and interference paths.}\label{path analysis}
\end{figure}

We design a compact backscatter tag using off-the-shelf components, enabling users to perform silent articulation in front of the tag at a short distance (Fig.~\ref{path analysis}). This tag-based design brings three practical benefits. 
First, the tag does not actively transmit RF signals; instead, it reflects and modulates an external RF source, which allows extremely low-power operation and makes backscatter suitable for battery-free or ultra-low-power edge sensing~\cite{xiao2020motion}. 
Second, by introducing a controlled frequency shift, the tag separates the sensing-related backscatter component from strong direct-path Wi-Fi signals, facilitating reliable extraction under co-channel conditions. 
Third, backscatter tags are small, lightweight, and low-cost, making them easy to deploy in everyday settings. Unlike prior RF lip-reading systems that require attaching or holding a tag near the chin~\cite{wang2019rfid,zhang2022hearme}, {\systemname} allows users to place the tag on nearby objects (e.g., a stand or monitor) and perform non-contact silent input simply by facing the tag.

We implement frequency-shifted backscatter by periodically switching the tag impedance with a known modulation $m(t)$ whose fundamental frequency is $\Delta f_1$ (e.g., a square-wave switching). Let $x(t)$ denote the complex baseband Wi-Fi waveform emitted by the RF source (e.g., an OFDM signal). The received baseband signal can be written as
\begin{equation}\label{eq:backscatter_mix}
r(t)=h_d(t)\,x(t) \;+\; h_b(t)\,m(t)\,x(t) \;+\; n(t),
\end{equation}
where $h_d(t)$ denotes the direct (unmodulated) path channel, $h_b(t)$ denotes the tag-reflected path channel, and $n(t)$ is noise. The multiplication by $m(t)$ shifts the spectrum of $x(t)$ to multiple sidebands around $\pm k\Delta f_1$. In practice, we isolate the first-order shifted component (around $\pm\Delta f_1$) via band-pass filtering (or equivalently, digital downconversion by $\Delta f_1$ followed by low-pass filtering). This yields a clean backscatter component whose amplitude/phase variations are dominated by $h_b(t)$, on which we subsequently extract lip-motion features.

\subsubsection{Motion Interference Suppression for Lip-motion Traces}
After isolating the tag-related backscatter component, we further suppress motion interference to obtain a clean lip-motion signal. In practice, interference mainly comes from motions of other people and objects in the same scene and from the user’s own non-lip motions (e.g., slight head sway, breathing, and blinking). 
Let $\tilde r(t)$ denote the complex baseband signal of the first-order frequency-shifted backscatter component after band-pass filtering. We compute an instantaneous phase-difference sequence:
\begin{equation}\label{eq:phase_diff}
\Delta\phi(t)=\angle\big(\tilde r(t)\tilde r^{*}(t-\Delta t)\big),
\end{equation}
where $(\cdot)^*$ denotes complex conjugate and $\Delta t$ is the sampling interval. This phase-difference sequence approximates instantaneous frequency (Doppler-related) variations and largely removes constant phase offsets and slow phase drifts. To suppress high-frequency measurement noise, we apply light smoothing and a low-pass filter (0--50~Hz) to $\Delta\phi(t)$, retaining the dominant dynamics of articulatory motions.

External motions from other people/objects can still introduce bursty disturbances even after isolating the shifted backscatter component. 
We therefore apply a simple yet reproducible gating strategy on the filtered $\Delta\phi(t)$ using a sliding window. 
For each window, we compute the fluctuation magnitude $g(t)$ as the root-mean-square (RMS) value of $\Delta\phi(t)$. 
A window is marked as contaminated if $g(t)$ is significantly larger than the typical level, i.e., $g(t)>\mathrm{median}(g)+\alpha\cdot\mathrm{MAD}(g)$, where $\mathrm{MAD}(\cdot)$ is the median absolute deviation and $\alpha$ is a constant (set to 3 in our implementation). 
We then discard the contaminated windows and keep the remaining segments for subsequent processing, which reduces the impact of strong non-target movements in the environment.

The remaining interference is mainly caused by the user’s involuntary head/body sway and subtle facial motions such as breathing and blinking, which can overlap with lip motions in time. To separate these components, we adopt variational mode decomposition (VMD)~\cite{dragomiretskiy2013variational}, which decomposes a non-stationary signal into a small set of narrow-band modes with different center frequencies. Given the filtered phase-derived trace, VMD estimates $K$ modes $\{u_k(t)\}$ and their center frequencies $\{\omega_k\}$ by solving a constrained variational optimization problem (following the standard formulation in~\cite{dragomiretskiy2013variational}). We set $K$ to a small value (e.g., $K=4$ in our implementation) and retain the mode(s) whose center frequencies fall into the lip-motion band, while suppressing slower modes dominated by breathing, blinking and head sway. Finally, we reconstruct the lip-motion signal by summing the retained modes, producing a stable input for subsequent feature extraction and lexicon-guided decoding.

\subsection{Feature Extraction of Lip-reading Units}
\label{sec:Feature Extraction} 

\subsubsection{Segmenting Lip-motion Units}
Similar to spoken language, continuous silent articulation can be viewed as a composition of short, recurring motion patterns. We refer to these repeatable segments in wireless lip-motion signals as \emph{lip-motion units}. They serve as syllable-like building blocks in the sensing domain and can be concatenated to form complete sentences, while not necessarily aligning with linguistic syllable boundaries. Modeling at the unit level helps reduce sequence complexity and improves the robustness of downstream decoding, especially when phoneme-level cues are weak or ambiguous in wireless observations.
Segmenting lip-motion units from RF signals is challenging. First, unit-level segmentation for wireless lip sensing has been rarely studied, and visual boundary cues cannot be directly transferred to RF measurements. Even with video as an auxiliary reference, annotating unit boundaries precisely is labor-intensive and difficult to scale. Second, unit duration and boundary locations vary across speakers and articulation rates, making fixed templates unreliable.

\begin{algorithm}[t]
\caption{Lip-motion Unit Segmentation}
\label{alg1}
\LinesNumbered
\DontPrintSemicolon
\KwIn{Signal $S$; window sizes $w_s$ (short), $w_l$ (long), $w_f$ (fine); thresholds $\lambda_1,\lambda_2$}
\KwOut{Candidate unit segments $\mathcal{I}$}
Compute short-term energy $E_s(t)$ using window $w_s$ and long-term baseline energy $E_l(t)$ using window $w_l$\;
Obtain an activity mask $A(t)$ by comparing $E_s(t)$ and $E_l(t)$ (set $A(t)=1$ if $E_s(t)>E_l(t)$, otherwise $A(t)=0$), and merge consecutive active indices into coarse regions $\{S_k\}$;
$\mathcal{I}\leftarrow \emptyset$\;
\ForEach{coarse region $S_k$}{
  Compute a fine-scale STE curve $STE(i)$ on $S_k$ using window $w_f$\;
  Select prominent peaks $\{(p_j,\tau_j)\}$ of $STE$ with $p_j>\lambda_1\cdot \max(STE)$\;
  Set a split threshold $\delta \leftarrow \lambda_2\cdot(\max(STE)-\min(STE))$\;

  \ForEach{adjacent peak pair $(p_j,\tau_j),(p_{j+1},\tau_{j+1})$}{
    \If{$|p_{j+1}-p_j|>\delta$}{
      Split at the valley $v=\arg\min_{\tau_j\le i\le \tau_{j+1}} STE(i)$\;
    }
  }
  Add the resulting sub-segments to $\mathcal{I}$\;
}
\textbf{return} $\mathcal{I}$\;
\end{algorithm}

To address these challenges, we design a lightweight algorithm that approximately segments candidate units based on energy dynamics, and then clusters similar segments into unit categories. As shown in Algorithm~\ref{alg1}, we first perform a coarse search using two sliding windows (0.2~s and 5~s) to detect short-term energy increases relative to a long-term baseline, producing candidate regions likely containing lip-motion activity. We then refine boundaries within each candidate region by computing a short-term energy (STE) trace with a finer window (0.1~s) and locating local peaks and troughs on the smoothed STE curve. A new unit boundary is inserted when the STE change between consecutive peaks exceeds a threshold $\delta$, which indicates a significant transition in the articulation pattern. We set $\delta$ adaptively according to the local energy range, $\delta=\lambda_2\cdot(p_{\max}-p_{\min})$, where $p_{\max}$ and $p_{\min}$ are the maximum and minimum STE values within the current candidate region. We use $\lambda_1=0.1$ and $\lambda_2=0.3$ empirically in our implementation.

\subsubsection{Clustering Similar Lip-reading Units}
Guided by the preliminary study, we cluster segmented lip-motion units based on features that capture both temporal energy dynamics and frequency-domain signatures of articulatory motion. Specifically, for each segmented unit, we use the filtered phase-difference sequence $\Delta\phi(t)$ as a 1-D motion representation, and compute features in both time and frequency domains.

For time-domain characteristics, we first derive a short-time energy (STE) curve from the motion trace and extract descriptive statistics to summarize its distribution, including maximum, minimum, mean, and quartiles, as well as standard deviation, skewness, and kurtosis. 
To capture finer variations, we further split each unit segment into 25 equal-duration sub-segments and compute the mean energy of the motion trace in each sub-segment. 
For frequency-domain characteristics, we compute the spectrum of the motion trace and extract standard spectral descriptors, including spectral centroid (the center of mass of spectral energy), spectral spread and spectral entropy (dispersion and concentration of energy around the centroid), and spectral skewness (asymmetry and peakedness of the spectral shape). In addition, to emphasize dynamic transitions within a unit, we apply first- and second-order differencing to the motion trace and re-extract the same set of time- and frequency-domain descriptors on the differenced signals, which better captures changes in articulation speed and acceleration.
In total, each unit segment is represented by a 67-dimensional feature vector. Before clustering, we standardize each feature dimension (z-score normalization) to mitigate scale differences across features. Finally, we apply K-means clustering on the resulting feature vectors and treat each cluster as a category of approximately similar lip-motion units, which serves as the minimum unit for subsequent modeling and decoding.

\subsubsection{Self-supervised Pre-training with Cluster-based Pseudo Labels}
Lip-motion patterns exhibit substantial variability across users due to coarticulation and context effects. As a result, unit segments that belong to the same cluster may still show non-trivial intra-class variations. To learn more robust and transferable representations before semantic decoding, we adopt a self-supervised pretraining scheme based on cluster-derived pseudo labels.
Specifically, after clustering segmented lip-motion units (Section~\ref{sec:Feature Extraction}), we treat each cluster ID as a pseudo label and train an encoder to predict the cluster assignment. For each unit segment, we compute a time--frequency representation using STFT (i.e., a spectrogram) to expose both local temporal dynamics and frequency-domain signatures of articulation. The resulting feature sequence is first linearly projected by a fully connected layer and then encoded by a Transformer encoder (12 layers, 12 attention heads per layer). A linear classifier followed by a softmax layer is used to predict the pseudo label, and the model is trained with a standard cross-entropy objective.
After pretraining, we reuse the Transformer encoder as an initialization for the downstream semantic inference model. This pretraining objective encourages the encoder to capture shared structure across similar units while being tolerant to within-cluster variations, which improves adaptation when learning context-dependent decoding in the next stage.

\begin{figure}[t]
  \centering
  \includegraphics[width=3.4in]{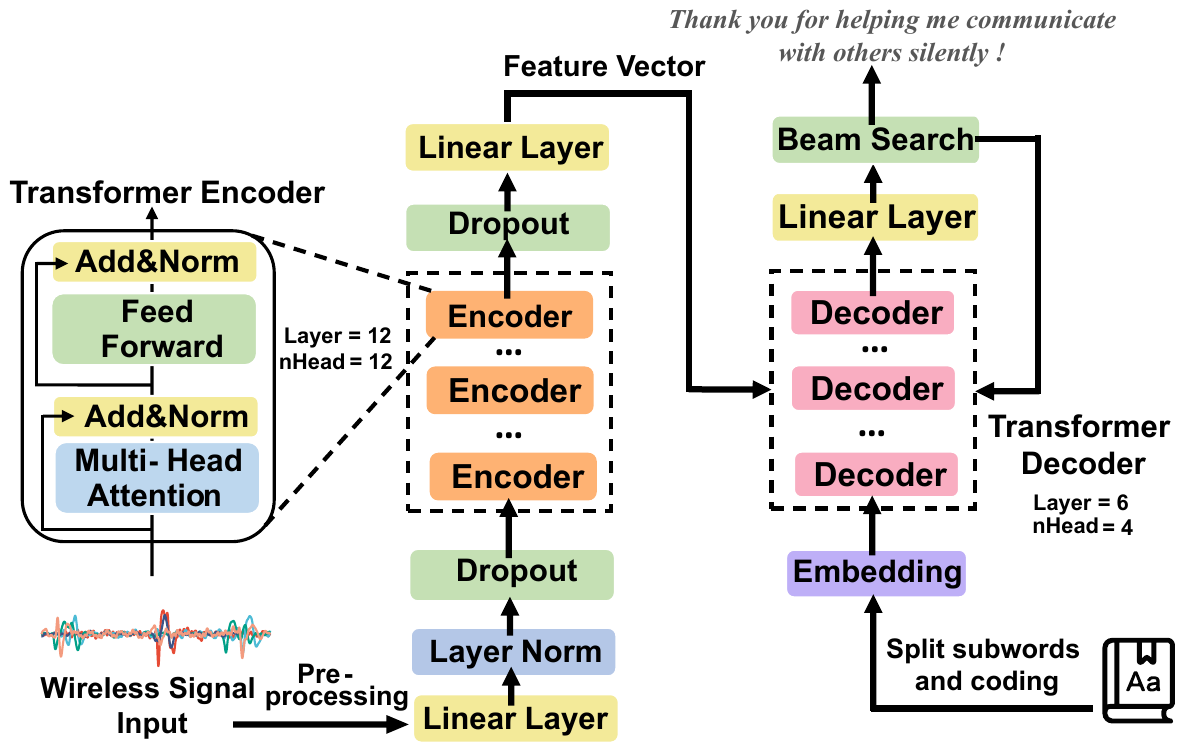} 
  \caption{Lexicon-guided Transformer encoder--decoder with beam search for open-sentence lip-motion decoding.}\label{Model architecture}
  \vspace{-0.05in}
\end{figure}

\subsection{Open-sentence Inference via Lexicon-based Decoding}\label{sec:Semantic Reasoning}

\subsubsection{Building the Lexicon}
{\systemname} allows users to initialize a task-specific lexicon from high-frequency words in a target scenario. Once the lexicon is defined, the system can recognize arbitrary sentences composed of words covered by the lexicon, which is important for practical silent interaction. For example, a smart-home ``silent assistant'' can prioritize frequently used command words rather than an open-domain vocabulary, improving accuracy under a fixed model capacity.

Given a scenario corpus, we first compute word frequencies $\{(\text{word}_i,\text{freq}_i)\}$. Each word is represented as a sequence of characters, and we append an end-of-word symbol \texttt{</w>} to the final character to distinguish word-final occurrences. For instance, \texttt{LOVE} is initialized as (\texttt{L}, \texttt{O}, \texttt{V}, \texttt{E</w>}). We then initialize the lexicon with all characters appearing in the corpus (including character+\texttt{</w>} forms). Next, we iteratively count frequencies of adjacent symbol pairs in the corpus and merge the most frequent pair into a new subword symbol, updating the corpus representation accordingly. This merge operation is repeated until the lexicon reaches a predefined size or the highest remaining pair frequency falls to 1.

A subword lexicon is beneficial because it naturally shares structure across morphological variants. For example, \texttt{looks}, \texttt{looking}, and \texttt{looked} can be decomposed into \texttt{look}+\texttt{s}, \texttt{look}+\texttt{ing}, and \texttt{look}+\texttt{ed}, respectively, instead of being treated as unrelated whole-word tokens. In our decoding, the model first identifies the stem-related subword (e.g., \texttt{look}) and then predicts the following subword(s) (e.g., \texttt{s}/\texttt{ing}/\texttt{ed}), enabling flexible sequence inference at the subword level.


\subsubsection{Lexicon-guided Subword Decoding}
We design a sequence-to-sequence lip-reading model with a Transformer-based encoder--decoder architecture (Fig.~\ref{Model architecture}). Following the unit pretraining stage, we convert each continuous lip-motion signal into a time--frequency representation using STFT, which captures both local temporal dynamics and Doppler-related spectral variations.

For efficiency, we downsample the input signal to 1~kHz and apply STFT with a 256-sample window (256~ms) and a hop size of 10 samples (10~ms). The resulting spectrogram forms a sequence of $T$ frames, where $T$ depends on the utterance duration; each frame is represented by $F$ frequency bins determined by the FFT setting. We pad sequences to a fixed length within each batch for efficient training.
The encoder first projects each frame-level input vector to a $D$-dimensional embedding ($D=768$) via a linear layer, followed by LayerNorm and dropout for regularization. A Transformer encoder (12 layers, 12 attention heads per layer) then produces contextualized representations $\mathbf{H}=[\mathbf{h}_1,\mathbf{h}_2,\ldots,\mathbf{h}_T]$, where each $\mathbf{h}_t \in \mathbb{R}^{D}$ summarizes the input context around frame $t$.
On top of the encoder, we employ a Transformer decoder (6 layers, 4 attention heads per layer) to generate the output sequence. The decoder uses cross-attention to attend to $\mathbf{H}$ and predicts a sequence of subword tokens from the lexicon (Section~\ref{sec:Semantic Reasoning}). We train the model end-to-end with a standard cross-entropy objective. 

At inference time, we adopt beam search to improve decoding accuracy. Specifically, at each decoding step we keep the top-$k$ partial hypotheses with the highest cumulative log-probabilities, expand them with candidate next tokens, and prune back to $k$ beams. Hypotheses that generate the end-of-sequence token (\texttt{<eos>}) are moved to a completed set, and decoding stops when all beams are completed. The final output is chosen as the completed hypothesis with the highest score. Since the output space is defined by the subword lexicon, adding new words that are composable from existing subwords does not require retraining model parameters.
Overall, {\systemname} converts continuous lip-motion traces into a subword-token sequence via lexicon-guided decoding, enabling open-sentence recognition within the lexicon. In the next section, we evaluate the end-to-end performance of {\systemname} and quantify the impact across users and scenarios.

\section{System Evaluation}\label{sec:Evaluation} 
\subsection{Experimental Deployment} 

\textbf{Implementation and Setup.} 
To obtain the raw signal data, we use LabVIEW to control a USRP-RIO transmitter and a USRP-2943R receiver to generate and capture Wi-Fi signals, following prior studies~\cite{hameed2022pushing,zhang2022hearme}. 
We adopt an Software-Defined Radio (SDR) prototype because it provides full access to PHY-layer signals and flexible control of the transmission/reception chain for controlled evaluation. 

\begin{figure}[h]
  \centering
  \includegraphics[width=3.5in]{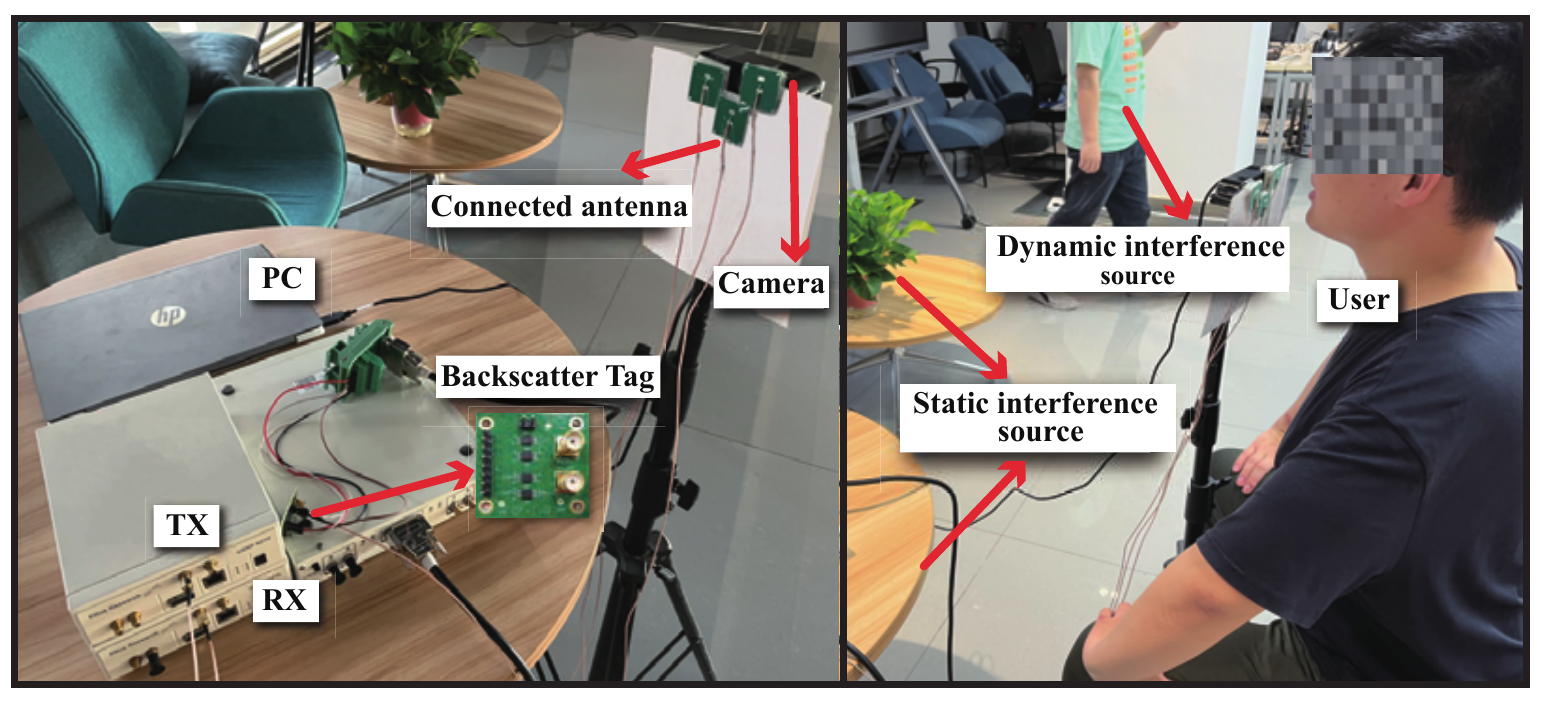} 
  \caption{Deployment and Prototype: prototype deployment in a real indoor environment, including the backscatter tag and antenna connected to TX/RX and PC, synchronized camera recording, and robustness evaluation under static and dynamic interference sources with a user in the loop.}\label{implementation}
  \vspace{-0.1in}
\end{figure}

Recent research has demonstrated that fine-grained CSI/CFR can be extracted on selected 802.11ax commercial Wi-Fi platforms via customized firmware/driver toolchains (e.g., AX-CSI on Broadcom 43684 devices and PicoScenes on Intel AX200/AX210 NICs), as well as toolkits that enable CSI collection from some commercial Wi-Fi APs \cite{gringoli2022ax,wang2025wi}. 
These advancements indicate that porting {\systemname} to commercial Wi-Fi hardware is theoretically completely feasible. However, customizing a complete COTS version is beyond the scope of this paper. Therefore, as in the previous Wi-Fi sensing research, we focus on implementing and deploying the SDR-based prototype system and evaluating its performance here.
To facilitate the user lip-reading tests, we fix the backscatter tags on a bracket during basic data collection and experiments, as shown in Fig.~\ref{implementation}. We also place a camera at the center to record video of users' lip movements for visual comparison.

\textbf{Data Collection.} 
We recruit 15 volunteers to conduct an extensive, real-world evaluation of {\systemname} over a six-month period.
The basic scene is an ordinary student activity space, including various furniture and clutter that create multipath disturbances.
In addition, we conducted multiple sets of experiments and evaluated 2 application cases in student offices, libraries, and dormitories. Volunteers lip-read within 50 cm of the backscatter tags and the system operating power was set to 20 dBm. To objectively evaluate {\systemname}'s lip-reading inference ability, instead of specifying simple sentences and words, we chose \textit{Everyday Conversation Handbook} \cite{Everyday}, published by the U.S. Department of State, to initialize the dictionary. It contains a total of 30 sections covering almost all the basic topics of everyday conversation. We organize it into 340 different sentences with a total of 3398 words. We ask all volunteers to first lip-read these sentences at least twice as the base data for training the model. Then, users could lip-read any sentence based on the words in constructed lexicon. We eventually collect 11,700 data samples in the base scenario.


\begin{figure*}[t]
  \centering
  \begin{minipage}[t]{0.50\textwidth}
    \centering
\includegraphics[width=\linewidth]{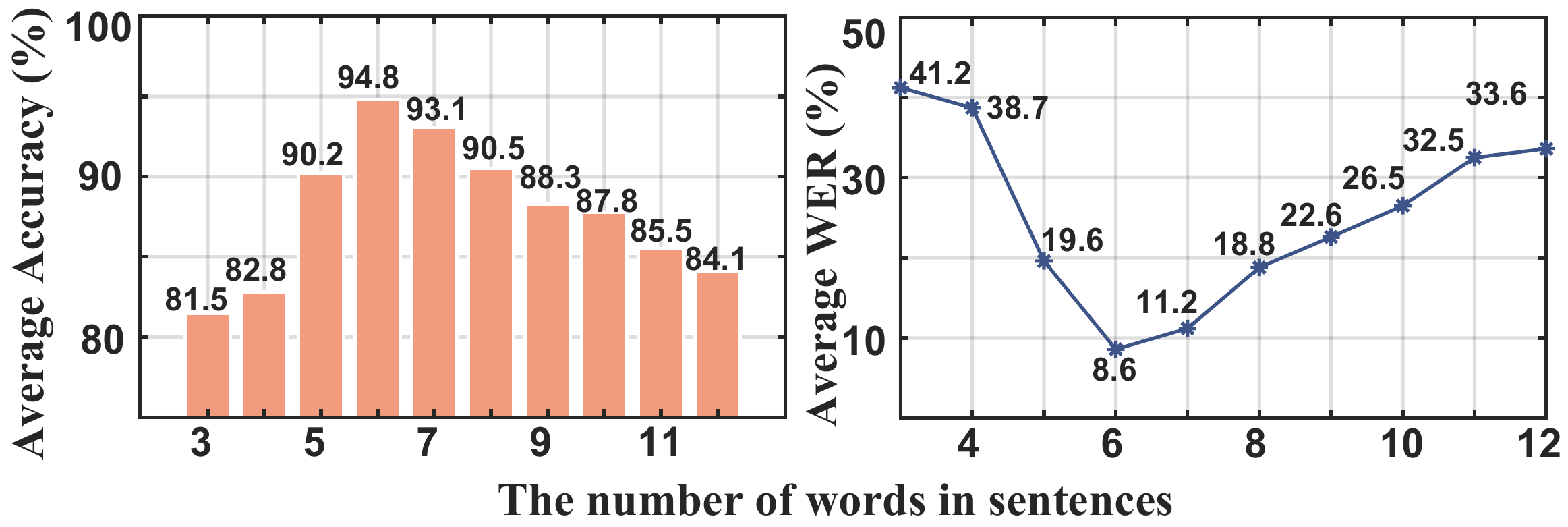}
    \captionof{figure}{Overall performance of {\systemname}.}
    \label{fig:Overall}
  \end{minipage}\hfill
  \begin{minipage}[t]{0.48\textwidth}
    \centering
    \begin{minipage}[t]{0.49\linewidth}
      \centering
\includegraphics[width=\linewidth]{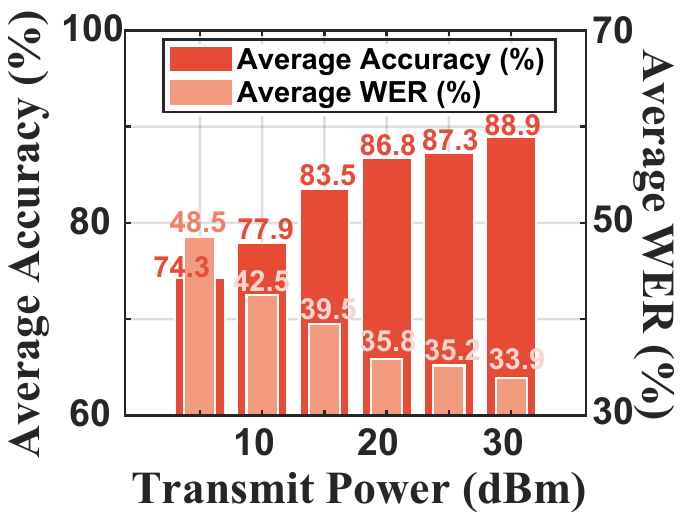}
    \end{minipage}\hfill
    \begin{minipage}[t]{0.49\linewidth}
      \centering
\includegraphics[width=\linewidth]{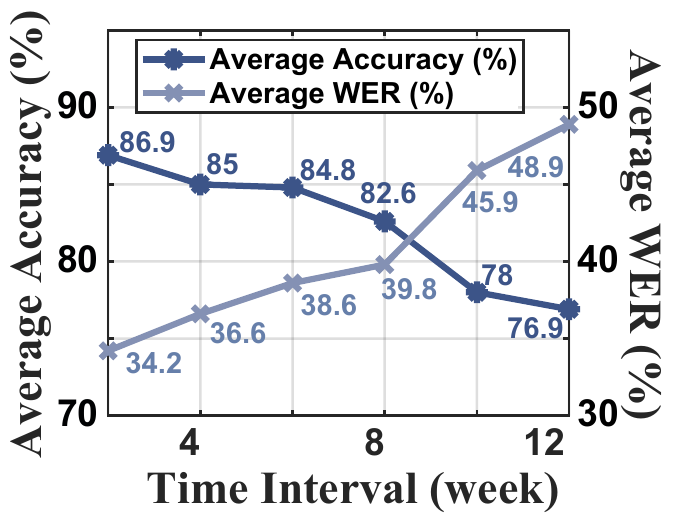}
    \end{minipage}
    \captionof{figure}{Impact of Signal Strength and Time Interval.}
    \label{fig:ImpactSandT}
  \end{minipage}
\end{figure*}

\begin{figure*}[t]
\begin{minipage}[t]{0.5\linewidth}
\centering
\includegraphics[height=1.3in]{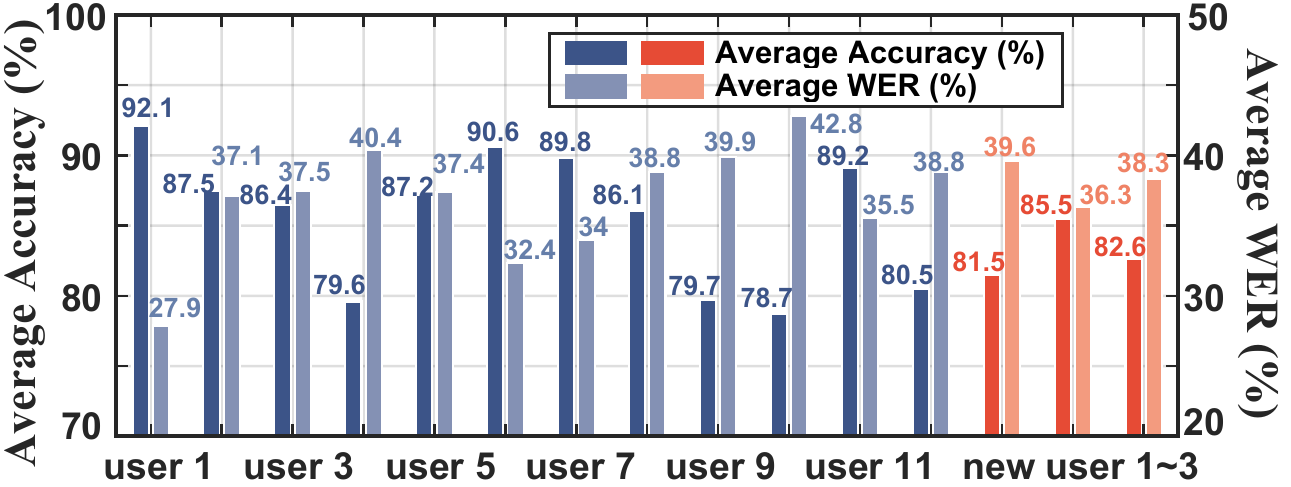}
\vspace{-0.05mm}
\caption{Experimental results of system generalization.}\label{fig:newuser}
\end{minipage}%
\begin{minipage}[t]{0.5\linewidth}
\centering
\includegraphics[height=1.3in]{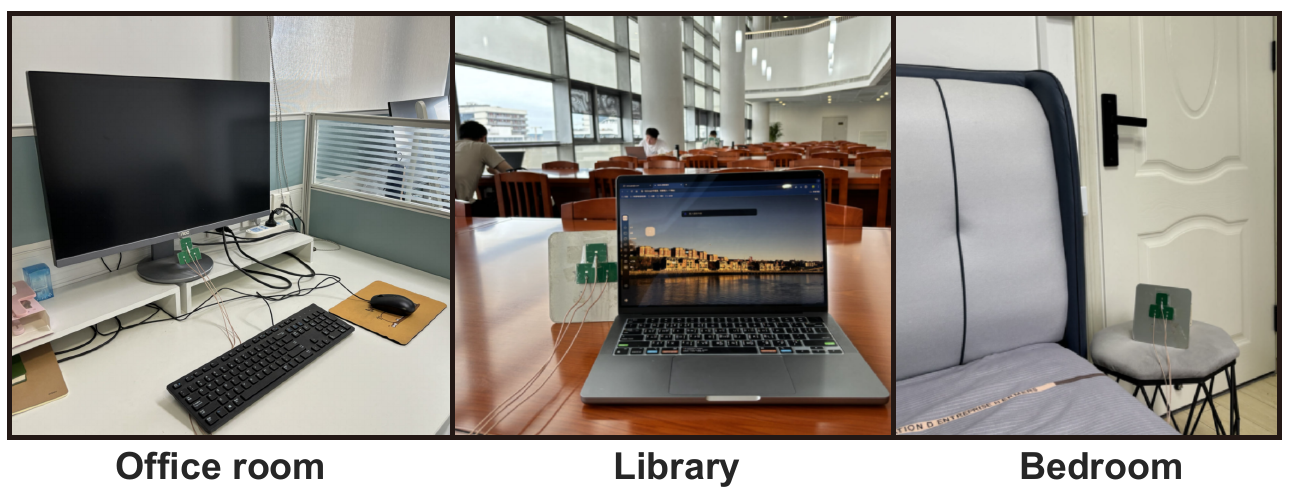}
\vspace{-0.05mm}
\caption{Experiments in real situations.}\label{fig:real situations}
\end{minipage}
\end{figure*}


\subsection{Overall Performance}  
\textbf{Evaluation Metrics.} 
We calculate accuracy and Word Error Rate (WER) as the primary evaluation metrics for assessing the {\systemname}'s performance. 
Accuracy indicates the proportion of words correctly predicted.
WER is a widely used evaluation metric in continuous speech recognition, calculated by Eq.(\ref{eq:WER}):

\begin{equation}\label{eq:WER}  
{\rm{WER}} = \frac{{{N_s} + {N_d} + {N_i}}}{{{N_s} + {N_d} + {N_c}}},
\end{equation}  
where ${N_c}$ represents the number of correctly identified words, and $N_s$, $N_d$ and $N_i$ denote the number of words replaced, deleted and inserted in the predicted result compared to the actual value. A lower value means that the predicted sentence is close to the real sentence and the performance is better. 


\textbf{Average Accuracy and WER.} We retain 3 users as new users to test the system's generalization to unknown users. For the remaining data, we randomly select 80\% for training and 20\% for testing. As shown in Fig. \ref{fig:Overall}, the results show that \systemname~achieves an average accuracy of 85.61\% for predicting words in sentences and an average WER of 36.87\% in complete sentences.
Meanwhile, we use a SOTA open-source lip-reading recognition framework \cite{ma2022visual} to process the recorded visual data, with an average WER of 32.3\%. Therefore, the overall results indicate that Lip-Siri performs comparably to representative visual-based SSI in the same test environment.


\subsection{Impact of Different Sentence Lengths}
To evaluate the performance on sentences of varying lengths, we group evaluations on sentences with a distribution of 3-12 words, as shown in Fig. \ref{fig:Overall}.
The result shows that the system performs best for sentences with 5-7 words, achieving an average accuracy of over 90\% and a WER below 20\%. Subsequently, the system's performance slightly declines as the number of words continues to increase. This is because the difficulty of learning alignment relationships increases significantly with more words in a sentence. Overall, the system's lip-reading comprehension capability is good. Even for sentences with 12 words, the word accuracy reaches 84.1\% and the WER is 33.6\%.


\subsection{Impact of Signal Strength}
To test the performance of the system in different Settings, we invite 3 volunteers to lip-read any 50 sentences at different  signal strengths. 
As shown in Fig.\ref{fig:ImpactSandT}, the results indicate that the system's recognition performance is poor when the transmitter power is set to 5 dBm. This is because the signal is too weak, affecting the data quality received.
With the transmitter-side signal transmission power increasing, the system performance improves. But beyond 20 dBm, the growth changes very slowly. Too high power will bring unnecessary energy consumption, so in our experiments, we set the transmitter-side signal to 20 dBm.

\subsection{Impact of Time Interval}
To test the stability of the system in the time dimension, we monitor the recognition results of lip-reading data over a continuous period of 3 months. Fig.\ref{fig:ImpactSandT} shows the course of the changes, with an interval of 2 weeks. 
The results showed that the average accuracy of the system for word recognition decreased by only 4.3\% and the average WER of sentences increased by only 5.6\% during the 2-month time interval. 
This shows that the user lip-reading features extracted by {\systemname} are very stable, and the system performance does not fluctuate dramatically with the increase of time.
However, after more than two months, the decline in system performance will be faster.
We can consider fine-tuning the model by introducing new user lip reading data periodically to ensure that its accuracy always remains within an acceptable range.

\subsection{System Generalization}\label{sec:User-independent}
To study the system's generalization capability, we test it using lip-reading data from three new users.
As shown in Fig.\ref{fig:newuser}, for the three new users, the average accuracy of word recognition ranged from 81.5\% to 85.5\%, and the average WER of sentences ranged from 36.3\% to 39.6\%. This result is in the normal range of intervals compared to users with training data, and even higher than some of the original users with poor performance. The extreme difference of the average accuracy of word recognition of {\systemname} for different users is 13.4\% and the extreme difference of the average WER of sentences is 14.92\%. 
This shows that {\systemname} well extracts the high dimensional features of the user's lip-reading actions and has a good tolerance for differences among different users. 
                                  
\subsection{System Delay}
We develop a graphical interface using LabVIEW to provide interaction with the user, who clicks the start button before lip-reading and the end button after lip-reading. The collected data was transferred via Ethernet to a laptop computer (PC) with Intel(R) Core(TM) i5-11500 and 32GB RAM for processing. The latency of the system is defined as the time difference between when the user clicks on end and when \systemname~displays the lip-reading result on the PC.
Since the feature extraction for clustering the basic lip-reading units is only deployed during the training phase, during actual operation, the system's latency is primarily determined by data transmission, signal processing, and the lip reading inference model.
Data are transmitted and saved in real time via Ethernet with an average delay of no more than 50ms.
The signal processing module is the most time-consuming part, with an average latency of 525 ms. The average reasoning time of the lip-reading reasoning model for 10s lip-reading signals is 142.5ms. Therefore, the average latency of {\systemname} is 717.5 ms on the currently configured device, showing that it can meet the requirement of real-time inference. 


\section{Case Studies: Silent Interaction Applications}
{\systemname} can be applied to numerous life scenarios. To evaluate its performance in real-world settings, we deploy it in three typical SSI requirement scenarios:  an office room, a library, and a bedroom, and we test two representative tasks in these scenarios: message sending and silent assistant.


\subsection{Message Sending} 
In some quiet environments such as libraries and bedrooms, users may feel it inconvenient to use voice input, since it will disturb others or roommates.
In addition, in some shared spaces like offices, users may prefer not to speak the content due to privacy concerns when sending messages.
To test whether {\systemname} can assist users in these environments, we ask 2 volunteers to attach backscatter tags to various nearby objects, as shown in Fig. \ref{implementation}. 
They then lip-read any 50 sentences as messages sent to the system for recognition.
The results indicate that {\systemname} achieves an average accuracy of 86.72\% and an average WER of 36.87\% in the three scenarios. 
This demonstrates that {\systemname} can be used in complex and extensive text input scenarios, demonstrating its broad application potential.


\subsection{Silent Assistant} 
Since the vocabulary of interaction commands is relatively fixed, we can update a specific dictionary to improve {\systemname}'s accuracy on specific tasks. We count 100 common words used for voice interaction commands and updated the lexicon. We then invite 2 volunteers to lip-read arbitrary commands made up of lexicon in each of three scenarios, such as ``Set an alarm for 8 am.''
The results show that {\systemname} achieves an average accuracy of 92\%, which exceeds its performance in the benchmark test. This is because the lexicon of interactive commands is smaller and the accuracy of model predictions is improved.
Furthermore, to test {\systemname}'s potential in noisy environments, we play audio as interference noise in the bedroom and ask volunteers to turn on Siri on iphones for comparison. The results show that in a quiet environment, Siri achieves 98\% accuracy in recognizing these commands. However, with increasing noise, its accuracy significantly drops. In a noisy environment of 50 decibels, Siri's average recognition accuracy decreases to 31\%, while {\systemname} maintains an accuracy of 90.5\%. This indicates that {\systemname} can effectively assist users in interacting with their devices in noisy real-life scenarios.




\section{Comparison and Discussion}\label{sec:comparison}

\subsection{Comparison with SOTA sensing-based SSIs}
Table~\ref{tab:comparison} summarizes representative SSIs that infer speech content from sensing signals. 
Existing systems typically rely on either (i) wearable/handheld sensors (e.g., smartphones, earphones, smartwatches, or strain sensors) or (ii) contactless acoustic/RF sensing, and all of them formulate recognition as a closed-set task over a pre-defined list of words, commands, or sentences. 
In contrast, {\systemname} makes two key advances: (1) it leverages Wi-Fi backscatter sensing without requiring users to wear or hold any device, and (2) it enables open-sentence recognition by decoding a variable-length token sequence over a lexicon rather than selecting from a pre-enumerated set. 
This design substantially expands the output space from a finite set of commands/sentences to combinatorially many sentence compositions within the lexicon, which is important for practical silent input where users may express previously undefined sentences. 
Since these prior SSIs do not release code and datasets, we report the best performance numbers as stated in their papers and provide a qualitative comparison and discussion. 
Overall, compared with prior SSIs, {\systemname} offers three main advancements:

\textbf{(1) Open-sentence recognition for practical silent interaction.}
Prior sensing-based SSIs formulate recognition as a closed-set problem, where the output is restricted to a small, pre-defined list of words or sentences.
In contrast, {\systemname} performs lexicon-guided sequence decoding: it constructs an extensible lexicon and decodes a variable-length token sequence, enabling recognition of previously unseen sentence combinations composed from the lexicon.
Compared with training on fixed sentence, {\systemname} expands the interaction space from a finite utterance set to combinatorially many sentence compositions within the lexicon, substantially improving the practicality of SSIs.
This design can scale more naturally to real-world silent interaction (e.g., message sending and silent assistant), where users need new sentence combinations.
\textbf{(2) Broader usage scenarios and better user experience.}
Many existing sensing-based SSIs rely on wearable or handheld devices (e.g., smartphones, earphones, smartwatches, or on-body sensors) to capture lip motions, which can be inconvenient in daily use and may reduce accessibility for users. 
This motivates exploring contactless or infrastructure-assisted paradigms that minimize user instrumentation.
Among wireless approaches, RFID Tattoo~\cite{wang2019rfid} and HearMe~\cite{zhang2022hearme} can recognize lip motions using RF signals, but they require the sensing tag to be placed very close to the user's face (e.g., attached to the body or held near the chin), which still imposes a non-trivial usage burden. 
In contrast, {\systemname} decouples the tag from the user: the backscatter tag can be placed on everyday objects in the environment (e.g., a desk stand, monitor, or bedside fixture), and users only need to face the tag to perform silent input. 
By avoiding any on-body attachment or hand-held operation, {\systemname} enables a more natural and user-friendly interaction experience and supports a wider range of practical scenarios.
\textbf{(3) Low-power operation and affordable deployment.}
Many prior SSIs rely on \emph{active} sensing: wearable or handheld devices continuously emit probing signals (e.g., ultrasound or RF) and record echoes to track lip motions. Continuous transmission and high sampling rates can impose non-trivial energy and runtime overhead on battery-powered edge devices, which limits long-term or always-on usage.
In contrast, {\systemname} adopts a passive backscatter design: the tag does not actively transmit; it only reflects and modulates ambient Wi-Fi signals, while the receiver analyzes the reflected signals to infer lip-reading. As a result, the sensing front-end can operate with very low power consumption, making it more suitable for practical, prolonged silent-interaction scenarios.
Beyond power, {\systemname} also lowers the deployment barrier. High-frequency radars (e.g., mmWave and 4D radar) can provide higher sensing resolution, but their dedicated hardware is typically expensive, which limits ubiquitous and large-scale deployment. In contrast, {\systemname} leverages widely available Wi-Fi infrastructure in indoor environments and uses inexpensive backscatter tags (typically costing only a few dollars), enabling a lightweight and affordable deployment for everyday settings.

\begin{table*}[t]
  \centering
  \scalebox{0.85}{
    \begin{threeparttable}
      \begin{tabular}{llllll}
        \hline
        \textbf{} & \textbf{Sensing Signal} & \textbf{Wear or hold devices} & \textbf{DataSet} & \textbf{Performance} & \textbf{Recognition Scope} \\ \hline
        SoundLip \cite{zhang2021soundlip}  & Acoustic & Smartphone & 20 words and 70 sentences & 91.2\% accuracy & Closed-set (pre-defined sentence) \\
        LipReader  \cite{yin2023acoustic} & Acoustic & Smartphone & 50 words & 91.58\% accuracy & Closed-set (pre-defined word) \\
        EchoLip  \cite{akbar2025echolip} & Acoustic & Smartphone & 500 sentences & 13.9\% WER & Closed-set (pre-defined sentence) \\
        CELIP \cite{zhang2021celip} & Acoustic & VR headsets & 70 sentences & 90.8\% accuracy & Closed-set (pre-defined sentence) \\
        EarSSR  \cite{sun2024earssr} & Acoustic & Earphone & 50 words and phrase & 93\% accuracy & Closed-set (pre-defined word) \\
        Lipwatch  \cite{zhang2024lipwatch} & Acoustic \& IMU   & Smartwatch & 65 words and 80 commands & 13.7\% WER & Closed-set (pre-defined sentence) \\
        MuteIt \cite{srivastava2022muteit} & IMU & Earphone & 100 words & 91\% accuracy & Closed-set (pre-defined word) \\
        E-MASK \cite{kunimi2022mask} & Strain sensor & Mask with sensor & 21 commands & 84.4\% accuracy & Closed-set (pre-defined word) \\
        RFID Tattoo \cite{wang2019rfid} & Radio-frequency & RFID Tag & 100 words and 20 sentences & 86\% accuracy & Closed-set (pre-defined sentence) \\
        HearMe \cite{zhang2022hearme} & Radio-frequency & RFID Tag & 20 words & 88\% accuracy & Closed-set (pre-defined word) \\
        TWLip-Seq \cite{zhu2025twlip} & 4-D Radar & None & 30 sentences & 4.55\% WER & Closed-set (pre-defined sentence) \\
        mSilent \cite{zeng2023msilent} & mmWave radar & None & 1000 sentences & 9.5\% WER & Closed-set (pre-defined sentence) \\ \midrule
        \textbf{Lip-Siri} & \textbf{Wi-Fi} & \textbf{None} & \textbf{3398 words and 340 sentences} & \textbf{85.61\% accuracy} & \textbf{Open-sentence (Lexicon-based contextual decoding)} \\  \bottomrule
      \end{tabular}
    \end{threeparttable}
  }
  \caption{Comparison of {\systemname} with SOTA sensing-based SSIs. (Open-sentence indicates decoding a variable-length token sequence over a lexicon rather than classifying among a pre-enumerated set of utterances.)}
  \label{tab:comparison}
\end{table*}

\subsection{Comparison with SOTA Visual-based SSIs}
\begin{figure}[h]
  \centering
  \includegraphics[width=2.7in]{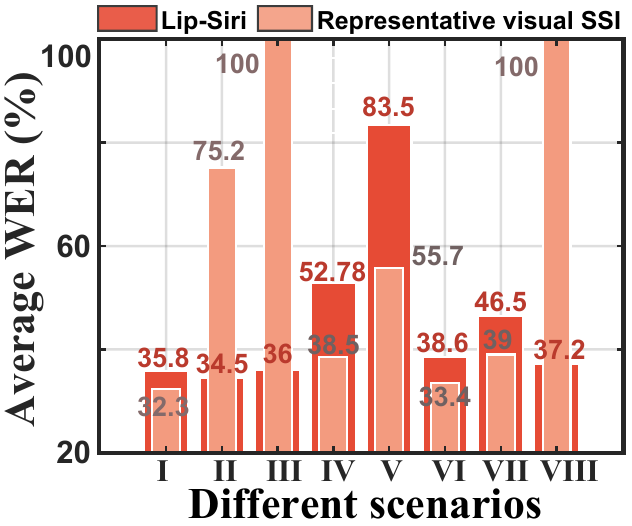}
  \caption{Comparison with SOTA Visual-based SSIs in 8 different actual scenarios: \Rmnum{1}. Normal, \Rmnum{2}. Low-light, \Rmnum{3}. Darkness, \Rmnum{4}. Head rotation, \Rmnum{5}. Jumping, \Rmnum{6}. Typing, \Rmnum{7}. Walking, \Rmnum{8}. Wear a mask.}
  \label{fig:sota-discussion}
\end{figure}

To benchmark {\systemname} against representative visual SSIs, we invited two volunteers to test 50 sentences in eight different scenarios and processed the recorded videos using an open-source SOTA visual SSI framework.
As shown in Fig.~\ref{fig:sota-discussion}, the visual SSI achieves an average WER of 32.3\% under normal conditions, slightly better than {\systemname}.
However, visual-based methods become unusable in the dark or when users wear masks, and their performance degrades substantially under poor lighting.
In contrast, {\systemname} maintains stable performance under these conditions.
We also observe that {\systemname}'s WER varies less under moderate body movements (e.g., head rotation, walking, and typing), indicating good tolerance to such motions.
However, {\systemname} may fail under violent movements (e.g., jumping), while the visual SSI remains relatively stable, highlighting complementary strengths across modalities.
This suggests that multi-modal fusion (e.g., combining visual and wireless cues) is a promising direction for building more robust SSIs, which we plan to explore in future work.

\newpage
\section{Conclusion}                   
In this paper, we present {\systemname}, a contactless silent speech interface that decodes silent lip motions using Wi-Fi backscatter sensing. {\systemname} supports open-sentence recognition via lexicon-guided subword decoding: users can initialize an extensible subword lexicon, and a Transformer encoder--decoder with beam search maps continuous lip-motion traces to variable-length sequences. Extensive experiments show that {\systemname} achieves 85.61\% accuracy on word prediction and a WER of 36.87\% on continuous sentence recognition. 
{\systemname} is fully contactless, enabling comfortable and privacy-friendly silent interaction, including when users wear masks. In future work, we will explore multi-modal fusion and broader real-world deployments across more diverse scenarios.

\section*{Conflict of Interest}
The authors declare that there are no conflicts of interest regarding the publication of this paper.

\bibliographystyle{IEEEtran}
\bibliography{references}

@article{samuel2024stroboscopy,
  title={Stroboscopy and acoustic analysis of voice following endotracheal intubation in otological surgeries},
  author={Samuel, John and others},
  journal={The Egyptian Journal of Otolaryngology},
  volume={40},
  number={1},
  pages={1--12},
  year={2024},
  publisher={Springer}
}

@article{aref2024vocal,
  title={Vocal dysfunction following thyroid surgery: a multidimensional subjective and objective study},
  author={Aref, Essam Eldin Mohamed and Ahmed, Gamal Abd El-Hamed and Ibrahim, Reham AbdEl-Wakil and Shrkawy, Aya Essam},
  journal={The Egyptian Journal of Otolaryngology},
  volume={40},
  number={1},
  pages={106},
  year={2024},
  publisher={Springer}
}

@article{wang2025role,
  title={The role of dynamic larynx CT in the investigation of persistent hoarseness following viral infections},
  author={Wang, Yong and Lin, Ziying and Zhu, Guanbin and Zhong, Hua and Xiao, Dongjian and Ma, Yanli and Zhuang, Peiyun and Cao, Dairong},
  journal={BMC Medical Imaging},
  volume={25},
  number={1},
  pages={441},
  year={2025},
  publisher={Springer}
}

@article{lakpa2025job,
  title={Job Accommodations and Job Loss From Unilateral Vocal Fold Paralysis},
  author={Lakpa, Koffi L and Bowen, Andrew and Menon, Ezra and Ring, Sydney and Nordby, Peter and Rasmussen, Miranda and Arroyo, Natalia and Zhao, Jiwei and Fernandes-Taylor, Sara and Francis, David O and others},
  journal={Otolaryngology--Head and Neck Surgery},
  volume={173},
  number={4},
  pages={938--946},
  year={2025},
  publisher={Wiley Online Library}
}

@article{akbar2025echolip,
  title={EchoLip: Pushing the Limit of Acoustic-Based Silent Speech Interface on Mobile Devices},
  author={Akbar, Ahsan Jamal and Guo, Kaiyi and Zhang, Qian and Wang, Dong},
  journal={IEEE Internet of Things Journal},
  year={2025},
  publisher={IEEE}
}

@article{zhu2025twlip,
  title={TWLip-Seq: A Novel Through-Wall Lip-Reading Approach via Sequence-to-Sequence Models and 4-D Radar},
  author={Zhu, Dongsheng and Han, Chong and Guo, Jian and Sun, Lijuan},
  journal={IEEE Sensors Journal},
  year={2025},
  publisher={IEEE}
}

@inproceedings{gringoli2022ax,
  title={AX-CSI: Enabling CSI extraction on commercial 802.11 ax Wi-Fi platforms},
  author={Gringoli, Francesco and Cominelli, Marco and Blanco, Alejandro and Widmer, Joerg},
  booktitle={Proceedings of the 15th ACM Workshop on Wireless Network Testbeds, Experimental Evaluation \& CHaracterization},
  pages={46--53},
  year={2022}
}

@article{wang2025wi,
  title={Wi-Fi Sensing Tool Release: Gathering 802.11 ax Channel State Information from a Commercial Wi-Fi Access Point},
  author={Wang, Zisheng and Li, Feng and Zhao, Hangbin and Mao, Zihuan and Zhang, Yaodong and Huang, Qisheng and Cao, Bo and Cao, Mingming and He, Baolin and Hou, Qilin},
  journal={arXiv preprint arXiv:2506.16957},
  year={2025}
}

@article{sun2024earssr,
  title={EarSSR: Silent Speech Recognition via Earphones},
  author={Sun, Xue and Xiong, Jie and Feng, Chao and Li, Haoyu and Wu, Yuli and Fang, Dingyi and Chen, Xiaojiang},
  journal={IEEE Transactions on Mobile Computing},
  year={2024},
  publisher={IEEE}
}

@article{yin2023acoustic,
  title={Acoustic-based Lip Reading for Mobile Devices: Dataset, Benchmark and A Self Distillation-based Approach},
  author={Yin, Yafeng and Wang, Zheng and Xia, Kang and Xie, Lei and Lu, Sanglu},
  journal={IEEE Transactions on Mobile Computing},
  year={2024},
  publisher={IEEE}
}

@article{zhang2024lipwatch,
  title={Lipwatch: Enabling Silent Speech Recognition on Smartwatches using Acoustic Sensing},
  author={Zhang, Qian and Lan, Yubin and Guo, Kaiyi and Wang, Dong},
  journal={Proceedings of the ACM on Interactive, Mobile, Wearable and Ubiquitous Technologies},
  volume={8},
  number={2},
  pages={1--29},
  year={2024},
  publisher={ACM New York, NY, USA}
}

@article{dragomiretskiy2013variational,
  title={Variational mode decomposition},
  author={Dragomiretskiy, Konstantin and Zosso, Dominique},
  journal={IEEE transactions on signal processing},
  volume={62},
  number={3},
  pages={531--544},
  year={2013},
  publisher={IEEE}
}

@article{dong2023electromyogram,
  title={Electromyogram-Based Lip-Reading via Unobtrusive Dry Electrodes and Machine Learning Methods},
  author={Dong, Penghao and Song, Yuanqing and Yu, Shangyouqiao and Zhang, Zimeng and Mallipattu, Sandeep K and Djuri{\'c}, Petar M and Yao, Shanshan},
  journal={Small},
  volume={19},
  number={17},
  pages={2205058},
  year={2023},
  publisher={Wiley Online Library}
}

@article{eskes2017predicting,
  title={Predicting 3D lip shapes using facial surface EMG},
  author={Eskes, Merijn and van Alphen, Maarten JA and Balm, Alfons JM and Smeele, Ludi E and Brandsma, Dieta and van der Heijden, Ferdinand},
  journal={PLoS One},
  volume={12},
  number={4},
  pages={e0175025},
  year={2017},
  publisher={Public Library of Science San Francisco, CA USA}
}

@article{janke2017emg,
  title={EMG-to-speech: Direct generation of speech from facial electromyographic signals},
  author={Janke, Matthias and Diener, Lorenz},
  journal={IEEE/ACM Transactions on Audio, Speech, and Language Processing},
  volume={25},
  number={12},
  pages={2375--2385},
  year={2017},
  publisher={IEEE}
}

@inproceedings{son2017lip,
  title={Lip reading sentences in the wild},
  author={Son Chung, Joon and Senior, Andrew and Vinyals, Oriol and Zisserman, Andrew},
  booktitle={Proceedings of the IEEE conference on computer vision and pattern recognition},
  pages={6447--6456},
  year={2017}
}

@inproceedings{chung2017lip,
  title={Lip reading in the wild},
  author={Chung, Joon Son and Zisserman, Andrew},
  booktitle={Computer Vision--ACCV 2016: 13th Asian Conference on Computer Vision, Taipei, Taiwan, November 20-24, 2016, Revised Selected Papers, Part II 13},
  pages={87--103},
  year={2017},
  organization={Springer}
}

@inproceedings{zhang2023echospeech,
  title={EchoSpeech: Continuous Silent Speech Recognition on Minimally-obtrusive Eyewear Powered by Acoustic Sensing},
  author={Zhang, Ruidong and Li, Ke and Hao, Yihong and Wang, Yufan and Lai, Zhengnan and Guimbreti{\`e}re, Fran{\c{c}}ois and Zhang, Cheng},
  booktitle={Proceedings of the 2023 CHI Conference on Human Factors in Computing Systems},
  pages={1--18},
  year={2023}
}

@inproceedings{wark1998approach,
  title={An approach to statistical lip modelling for speaker identification via chromatic feature extraction},
  author={Wark, Tim and Sridharan, Sridha and Chandran, Vinod},
  booktitle={Proceedings. Fourteenth International Conference on Pattern Recognition (Cat. No. 98EX170)},
  volume={1},
  pages={123--125},
  year={1998},
  organization={IEEE}
}

@inproceedings{skodras2011unconstrained,
  title={An unconstrained method for lip detection in color images},
  author={Skodras, Evangelos and Fakotakis, Nikolaos},
  booktitle={2011 IEEE International Conference on Acoustics, Speech and Signal Processing (ICASSP)},
  pages={1013--1016},
  year={2011},
  organization={IEEE}
}

@inproceedings{fan2012system,
  title={The system of face detection based on OpenCV},
  author={Fan, Xianghua and Zhang, Fuyou and Wang, Haixia and Lu, Xiao},
  booktitle={2012 24th Chinese control and decision conference (CCDC)},
  pages={648--651},
  year={2012},
  organization={IEEE}
}

@article{puviarasan2011lip,
  title={Lip reading of hearing impaired persons using HMM},
  author={Puviarasan, N and Palanivel, Sengottayan},
  journal={Expert Systems with Applications},
  volume={38},
  number={4},
  pages={4477--4481},
  year={2011},
  publisher={Elsevier}
}

@inproceedings{kunimi2022mask,
  title={E-MASK: a mask-shaped interface for silent speech interaction with flexible strain sensors},
  author={Kunimi, Yusuke and Ogata, Masa and Hiraki, Hirotaka and Itagaki, Motoshi and Kanazawa, Shusuke and Mochimaru, Masaaki},
  booktitle={Proceedings of the Augmented Humans International Conference 2022},
  pages={26--34},
  year={2022}
}

@article{zeng2023msilent,
  title={mSilent: Towards general corpus silent speech recognition using COTS mmWave radar},
  author={Zeng, Shang and Wan, Haoran and Shi, Shuyu and Wang, Wei},
  journal={Proceedings of the ACM on Interactive, Mobile, Wearable and Ubiquitous Technologies},
  volume={7},
  number={1},
  pages={1--28},
  year={2023},
  publisher={ACM New York, NY, USA}
}

@inproceedings{kellogg2016passive,
  title={Passive $\{$Wi-Fi$\}$: Bringing Low Power to $\{$Wi-Fi$\}$ Transmissions},
  author={Kellogg, Bryce and Talla, Vamsi and Gollakota, Shyamnath and Smith, Joshua R},
  booktitle={13th USENIX Symposium on Networked Systems Design and Implementation (NSDI 16)},
  pages={151--164},
  year={2016}
}

@article{xiao2020motion,
  title = {Motion-Fi{$^{+}$}: Recognizing and Counting Repetitive Motions With Wireless Backscattering},
  author={Xiao, Ning and Yang, Panlong and Yan, Yubo and Zhou, Hao and Li, Xiang-Yang and Du, Haohua},
  journal={IEEE Transactions on Mobile Computing},
  volume={20},
  number={5},
  pages={1862--1876},
  year={2020},
  publisher={IEEE}
}

@inproceedings{potamianos1997speaker,
  title={Speaker independent audio-visual database for bimodal ASR},
  author={Potamianos, Gerasimos and Cosatto, Eric and Graf, Hans Peter and Roe, David B},
  booktitle={Audio-Visual Speech Processing: Computational \& Cognitive Science Approaches},
  year={1997}
}

@inproceedings{chu2000bimodal,
  title={Bimodal speech recognition using coupled hidden Markov models},
  author={Chu, Stephen M and Huang, Thomas S},
  booktitle={Sixth International Conference on Spoken Language Processing},
  year={2000}
}

@article{assael2016lipnet,
  title={Lipnet: End-to-end sentence-level lipreading},
  author={Assael, Yannis M and Shillingford, Brendan and Whiteson, Shimon and De Freitas, Nando},
  journal={arXiv preprint arXiv:1611.01599},
  year={2016}
}

@article{shi2022learning,
  title={Learning lip-based audio-visual speaker embeddings with av-hubert},
  author={Shi, Bowen and Mohamed, Abdelrahman and Hsu, Wei-Ning},
  journal={arXiv preprint arXiv:2205.07180},
  year={2022}
}

@article{sugie1985speech,
  title={A speech prosthesis employing a speech synthesizer-vowel discrimination from perioral muscle activities and vowel production},
  author={Sugie, Noboru and Tsunoda, Koichi},
  journal={IEEE Transactions on Biomedical Engineering},
  number={7},
  pages={485--490},
  year={1985},
  publisher={IEEE}
}

@inproceedings{kapur2018alterego,
  title={Alterego: A personalized wearable silent speech interface},
  author={Kapur, Arnav and Kapur, Shreyas and Maes, Pattie},
  booktitle={23rd International conference on intelligent user interfaces},
  pages={43--53},
  year={2018}
}

@inproceedings{kimura2022silentspeller,
  title={SilentSpeller: Towards mobile, hands-free, silent speech text entry using electropalatography},
  author={Kimura, Naoki and Gemicioglu, Tan and Womack, Jonathan and Li, Richard and Zhao, Yuhui and Bedri, Abdelkareem and Su, Zixiong and Olwal, Alex and Rekimoto, Jun and Starner, Thad},
  booktitle={Proceedings of the 2022 CHI Conference on Human Factors in Computing Systems},
  pages={1--19},
  year={2022}
}

@article{zhang2022hearme,
  title={Hearme: Accurate and real-time lip reading based on commercial rfid devices},
  author={Zhang, Shigeng and Ma, Zijing and Lu, Kaixuan and Liu, Xuan and Liu, Jia and Guo, Song and Zomaya, Albert Y and Zhang, Jian and Wang, Jianxin},
  journal={IEEE Transactions on Mobile Computing},
  year={2022},
  publisher={IEEE}
}

@inproceedings{maier2005session,
  title={Session independent non-audible speech recognition using surface electromyography},
  author={Maier-Hein, Lena and Metze, Florian and Schultz, Tanja and Waibel, Alex},
  booktitle={IEEE Workshop on Automatic Speech Recognition and Understanding, 2005.},
  pages={331--336},
  year={2005},
  organization={IEEE}
}

@inproceedings{tan2017silenttalk,
  title={SilentTalk: Lip reading through ultrasonic sensing on mobile phones},
  author={Tan, Jiayao and Nguyen, Cam-Tu and Wang, Xiaoliang},
  booktitle={IEEE INFOCOM 2017-IEEE Conference on Computer Communications},
  pages={1--9},
  year={2017},
  organization={IEEE}
}

@article{wang2019rfid,
  title={Rfid tattoo: A wireless platform for speech recognition},
  author={Wang, Jingxian and Pan, Chengfeng and Jin, Haojian and Singh, Vaibhav and Jain, Yash and Hong, Jason I and Majidi, Carmel and Kumar, Swarun},
  journal={Proceedings of the ACM on Interactive, Mobile, Wearable and Ubiquitous Technologies},
  volume={3},
  number={4},
  pages={1--24},
  year={2019},
  publisher={ACM New York, NY, USA}
}

@ARTICLE{wifihear,
  author={Wang, Guanhua and Zou, Yongpan and Zhou, Zimu and Wu, Kaishun and Ni, Lionel M.},
  journal={IEEE Transactions on Mobile Computing}, 
  title={We Can Hear You with Wi-Fi!}, 
  year={2016},
  volume={15},
  number={11},
  pages={2907-2920},
  doi={10.1109/TMC.2016.2517630}}

@article{hameed2022pushing,
  title={Pushing the limits of remote RF sensing by reading lips under the face mask},
  author={Hameed, Hira and Usman, Muhammad and Tahir, Ahsen and Hussain, Amir and Abbas, Hasan and Cui, Tie Jun and Imran, Muhammad Ali and Abbasi, Qammer H},
  journal={Nature Communications},
  volume={13},
  number={1},
  pages={5168},
  year={2022},
  publisher={Nature Publishing Group UK London}
}

@article{gao2020echowhisper,
  title={Echowhisper: Exploring an acoustic-based silent speech interface for smartphone users},
  author={Gao, Yang and Jin, Yincheng and Li, Jiyang and Choi, Seokmin and Jin, Zhanpeng},
  journal={Proceedings of the ACM on Interactive, Mobile, Wearable and Ubiquitous Technologies},
  volume={4},
  number={3},
  pages={1--27},
  year={2020},
  publisher={ACM New York, NY, USA}
}

@article{zhang2021soundlip,
  title={Soundlip: Enabling word and sentence-level lip interaction for smart devices},
  author={Zhang, Qian and Wang, Dong and Zhao, Run and Yu, Yinggang},
  journal={Proceedings of the ACM on Interactive, Mobile, Wearable and Ubiquitous Technologies},
  volume={5},
  number={1},
  pages={1--28},
  year={2021},
  publisher={ACM New York, NY, USA}
}

@article{zhang2020endophasia,
  title={Endophasia: Utilizing acoustic-based imaging for issuing contact-free silent speech commands},
  author={Zhang, Yongzhao and Huang, Wei-Hsiang and Yang, Chih-Yun and Wang, Wen-Ping and Chen, Yi-Chao and You, Chuang-Wen and Huang, Da-Yuan and Xue, Guangtao and Yu, Jiadi},
  journal={Proceedings of the ACM on Interactive, Mobile, Wearable and Ubiquitous Technologies},
  volume={4},
  number={1},
  pages={1--26},
  year={2020},
  publisher={ACM New York, NY, USA}
}

@article{jin2022earcommand,
  title={EarCommand: " Hearing" Your Silent Speech Commands In Ear},
  author={Jin, Yincheng and Gao, Yang and Xu, Xuhai and Choi, Seokmin and Li, Jiyang and Liu, Feng and Li, Zhengxiong and Jin, Zhanpeng},
  journal={Proceedings of the ACM on Interactive, Mobile, Wearable and Ubiquitous Technologies},
  volume={6},
  number={2},
  pages={1--28},
  year={2022},
  publisher={ACM New York, NY, USA}
}

@misc{Everyday,
  author = {The U.S. Department of State},
  title = {{Everyday Conversations: Learning American English}},
  howpublished = {https://americanenglish.state.gov/resources/everyday-conversations-learning-american-english},
  year = {2023}, 
  note = "[Online]"
}

@article{afouras2018deep,
  title={Deep audio-visual speech recognition},
  author={Afouras, Triantafyllos and Chung, Joon Son and Senior, Andrew and Vinyals, Oriol and Zisserman, Andrew},
  journal={IEEE transactions on pattern analysis and machine intelligence},
  volume={44},
  number={12},
  pages={8717--8727},
  year={2018},
  publisher={IEEE}
}

@article{ma2022visual,
  title={Visual speech recognition for multiple languages in the wild},
  author={Ma, Pingchuan and Petridis, Stavros and Pantic, Maja},
  journal={Nature Machine Intelligence},
  pages={1--10},
  year={2022},
  publisher={Nature Publishing Group UK London}
}

@inproceedings{petridis2018end,
  title={End-to-end audiovisual speech recognition},
  author={Petridis, Stavros and Stafylakis, Themos and Ma, Pingehuan and Cai, Feipeng and Tzimiropoulos, Georgios and Pantic, Maja},
  booktitle={2018 IEEE international conference on acoustics, speech and signal processing (ICASSP)},
  pages={6548--6552},
  year={2018},
  organization={IEEE}
}

@inproceedings{ma2023auto,
  title={Auto-avsr: Audio-visual speech recognition with automatic labels},
  author={Ma, Pingchuan and Haliassos, Alexandros and Fernandez-Lopez, Adriana and Chen, Honglie and Petridis, Stavros and Pantic, Maja},
  booktitle={ICASSP 2023-2023 IEEE International Conference on Acoustics, Speech and Signal Processing (ICASSP)},
  pages={1--5},
  year={2023},
  organization={IEEE}
}

@inproceedings{khanna2021jawsense,
  title={JawSense: recognizing unvoiced sound using a low-cost ear-worn system},
  author={Khanna, Prerna and Srivastava, Tanmay and Pan, Shijia and Jain, Shubham and Nguyen, Phuc},
  booktitle={Proceedings of the 22nd International Workshop on Mobile Computing Systems and Applications},
  pages={44--49},
  year={2021}
}

@inproceedings{zhang2021celip,
  title={CELIP: Ultrasonic-based Lip Reading with Channel Estimation Approach for Virtual Reality Systems},
  author={Zhang, Yongzhao and Chen, Yi-Chao and Wang, Haonan and Jin, Xingyu},
  booktitle={Adjunct Proceedings of the 2021 ACM International Joint Conference on Pervasive and Ubiquitous Computing and Proceedings of the 2021 ACM International Symposium on Wearable Computers},
  pages={580--585},
  year={2021}
}

@article{srivastava2022muteit,
  title={MuteIt: Jaw Motion Based Unvoiced Command Recognition Using Earable},
  author={Srivastava, Tanmay and Khanna, Prerna and Pan, Shijia and Nguyen, Phuc and Jain, Shubham},
  journal={Proceedings of the ACM on Interactive, Mobile, Wearable and Ubiquitous Technologies},
  volume={6},
  number={3},
  pages={1--26},
  year={2022},
  publisher={ACM New York, NY, USA}
}

\newpage
 

\vfill
\end{document}